\documentclass[11pt]{article}

\usepackage{indentfirst}

\usepackage{amsmath,amssymb}
\usepackage{comment}
\usepackage[square,sort,comma,numbers]{natbib}

\usepackage{appendix}
\usepackage{graphicx}
\usepackage{psfrag,rotating,multirow,setspace,colordvi}
\usepackage{dcolumn}
\usepackage{bm,bbm}

\usepackage{subcaption}

\usepackage[font=footnotesize,format=plain,labelfont=bf]{caption}

\usepackage[usenames,dvipsnames]{color}
\usepackage{tikz}
\usepackage{pgfplots}
\usepackage{pgfplotstable}
\usepgfplotslibrary{groupplots}
\usepackage{scalefnt}
\usepackage{microtype}
\usepackage{xfrac}
\usepackage{relsize}
\usepackage{threeparttable}
\usepackage{setspace}


\newcommand{\cD}{\mathcal{D}}

\newcommand{\pS}{{\partial S}}

\newcommand{\dgamma}{{\dot{\gamma}_s}}

\newcommand{\bI}{{\mathbf{I}}}
\newcommand{\bn}{{\mathbf{n}}}
\newcommand{\bu}{{\mathbf{u}}}

\newcommand{\bx}{{\mathbf{x}}}

\newcommand{\bgamma}{{\dot{\boldsymbol\gamma}}}

\newcommand{\btau}{{\boldsymbol\tau}}

\pgfplotscreateplotcyclelist{newcolors}{
{RYB1,every mark/.append style={fill=RYB1,mark size={2.5}},mark=*},
{RYB2,every mark/.append style={fill=RYB2},mark=square*},
{RYB3,every mark/.append style={fill=RYB3,mark size={3}},mark=triangle*},
{RYB4,every mark/.append style={fill=RYB4,mark size={4}},mark=x},
{RYB5,every mark/.append style={fill=RYB5,mark size={3}},mark=oplus},
{RYB7,every mark/.append style={fill=RYB7},mark=*},
}
\pgfplotsset{
    standard/.style={
    thick,
    compat=1.8,
            scale only axis,
        width=0.45\textwidth,
        enlarge x limits=0.05,
        enlarge y limits=0.05,
        max space between ticks=40,
	cycle list name=newcolors
    }
}

\usepackage[colorinlistoftodos, color=blue!20!white, bordercolor=gray,
  textsize=tiny,textwidth=0.8in]{todonotes}

\begin{document}

\begin{center}
\textbf{\Large The non-homogeneous flow of 
a thixotropic fluid around a sphere}
\bigskip

\textbf{Jaekwang Kim}$^1$,  and \textbf{Jun Dong Park}$^{2,*}$ \\
\bigskip
$^1$Department of Mechanical Science \& Engineering,\\

\textit{University of Illinois at Urbana--Champaign}, Urbana, Illinois, 61801\\
$^2$Department of Chemical Engineering,\\
\textit{Kumoh National Institute of Technology}, Gumi, Gyeongbuk, 39177\\ 

jk12@illinois.edu, jdp@kumoh.ac.kr\\
$^*$corresponding author 
\end{center}

\bigskip

\begin{center}

\textbf{Abstract }\\
\bigskip
\begin{minipage}{0.85\textwidth}
The non-homogeneous flow of a thixotropic fluid 
around a settling sphere is simulated.
A four-parameter Moore model is used 
for a generic thixotropic fluid 
and discontinuous Galerkin method 
is employed to solve the structure-kinetics equation
coupled with the conservation equations 
of mass and momentum. 
Depending on the normalized falling velocity $U^{*}$,
which compares the time scale of structure formation and destruction,
flow solutions are divided into 
three different regimes, which are attributed to an interplay
of three competing factors:
Brownian structure recovery, shear-induced structure breakdown, 
and the convection of microstructures.
At small $U^{*}( \ll 1)$, 
where the Brownian structure recovery is predominant, 
the thixotropic effect is negligible and flow solutions are not too dissimilar to 
that of a Newtonian fluid.
As $U^{*}$ increases, a remarkable structural gradient is observed
and the structure profile around the settling sphere is determined
by the balance of all three competing factors. 
For large enough $U^{*}(\gg 1)$, 
where the Brownian structure recovery becomes negligible, 
the balance between shear-induced structure breakdown 
and the convection plays a decisive role in determining 
flow profile. 
To quantify the interplay of three factors,
the drag coefficient Cs of the sphere is investigated for ranges of $U^{*}$. 
With this framework, the effect of the destruction parameter, 
the confinement ratio, and a possible nonlinearity in the model-form 
on the non-homogeneous flow of
a thixotropy fluid have been addressed.   
\end{minipage}
\end{center}


\section{Introduction}

Thixotropy is a distinct rheological phenomenon
that formally signifies
``the continuous decrease of viscosity 
with time when flow is applied to a material 
that has been previously at quiescent state, 
and the subsequent recovery of viscosity 
when flow ceases"~\cite{Mewis:2011}. 
Such time-history effect is to be differentiated 
from viscoelasticity in that the thixotropic material 
naturally recovers initial viscosity 
after cessation of flow~\cite{Larson:2019}. 
The origin of thixotropy is associated with 
the gradual breakdown and building up 
of the microstructure~\cite{Mewis:2006}.

Thixotropic materials are often encountered 
in industrial processes including mining, 
wastewater treatment, printing inks, 
oil pipeline-transport, consumer applications (e.g cosmetics and food)
as well as biological fluids such as blood 
\cite{Food1,thixo_example_1, thixo_example_2, thixo_example_3,thixo_example_4}.
Occasionally, thixotropy has been intentionally 
built into commercial products for the convenience of their uses.
For example, thixotropic behavior of printing 
and coating materials greatly facilitates 
whole processes from storage to application and drying. 
In the storage process, highly stable 
materials with large viscosity are required 
to prevent sedimentation problem. 
By contrast, the application process 
requires flowable materials with low viscosity at high shear rates. 
In the following drying process, 
rapid recovery of the viscosity is needed, 
which plays a crucial role in the control of 
leveling and sagging problems \cite{Armelin:2006}. 
Introducing thixotropy to such products makes 
it possible to satisfy these required conditions 
at each stage. 

The ubiquity and growing importance of thixotropy 
leads to considerable effort to understand thixotropy of materials.
Numerous studies have dealt with thixotropy with a variety of approaches.
For instance, step experiments, hysteresis loops, and shear startup 
experiments have been typically employed in experimental approaches to characterize thixotropic fluids
\cite{Barnes:1997}. 
What lies at the heart of these studies is the understanding of
the relationship between their rheology and
microstructural change~(e.g., size of the floc, alignment of fibers, 
spatial distribution, and entanglement density).
Theoretical studies of thixotropy 
aim to interpret and predict the thixotropic behaviors of material.
In this regard,
numerous models have been proposed for thixotropy in literatures.
Excellent reviews are available at other sources 
\cite{Barnes:1997,Mewis:2009}, 
but in general, the existing thixotropy models 
are categorized into 3 groups; 
structural kinetics models \cite{Goodeve,Moore:1959}, 
continuum mechanical models \cite{Stickel:2006,Goddard:1984} 
and micro-mechanical models \cite{Patel:1988,Potanin:1991}.  

Though previous studies have established 
well-grounded knowledge on thixotropy, 
practical application is often limited. 
Most experimental approaches on thixotropy
mainly analyze material behavior at 
a homogeneous flow field (e.g.~one-dimensional shear),
which is induced from 
a few of commercialized flow geometries at rheometers. 
Likewise, many of the theoretical modeling works
have been formulated under the assumption of homogeneous flow field as well. 
However, actual applications of thixotropic materials often include
non-homogeneous flow fields. 
For example, mixing of thixotropic fluid is performed 
in various batches with complicated geometries 
that are designed for performance improvement.
In the batches, a heterogeneous shear field develops at 
different positions due to geometrical hindrance
\cite{Dickey:2004,Metzner:1957}.

In the presence of such non-trivial geometrical factors, 
the material thixotropy shows 
not only time dependent but also spatial dependent behavior, 
because fluid elements accumulate
different shear history according to different pathlines.
Thus, previous studies can provide an insight solely on 
intrinsic thixotropy with some limitation on the 
knowledge on rheology in real flow scenarios.
Consequently, the lack of understanding on 
non-homogeneous thixotropic fluid flows
causes hardship for design and control of 
material process with thixotropy.

The specific aim of the present work is to obtain a basic idea 
for interpreting a non-homogeneous flow of thixotropic fluid. 
In a strict sense, many of materials of interest, 
such as bloods and drilling fluids, display behavior that is 
a combination of thixotropic, elastic and viscous properties. 
Therefore, it is more reasonable and realistic to adopt a sophisticated 
rheological model with viscoelasticity and time dependency. 
Such rheological behaviors may or may not be observed, 
depending on the magnitude of stress and 
the timescales of each properties 
compared to an observation timescale. 
In an effort to address this issue, 
a few studies have been conducted.
These studies include modeling of thixo-elasto-visco-plastic
(TEVP) fluid in circular contraction-expansion asymmetric pipe \cite{new_cite1,new_cite2,new_cite3,new_cite4}.
It is novel and meaningful that these studies have 
provided quantitative solution for non-homogenous flow of 
TEVP fluid in a non-trivial flow geometry, 
which is challenging in terms of both rheology and numerical analysis.   
Meanwhile, the use of the sophisticated model makes it hard 
to separate thixotropic effect and viscoelastic effect, 
which prevents intuitive understanding on the role of each property. 
By narrowing our attention to purely-viscous thixotropic fluids, 
we seek to understand the interplay
between thixotropy and fluid geometry.  
Such insight will be useful not only for understanding 
of more elaborated rheological behavior such as in TEVP fluid, 
but be readily extended to much more 
complicated flow scenarios in industrial material process.
This specific goal makes current study distinct from other studies. 

We consider a specific model problem, 
which is a steady thixotropic flow around a sedimenting sphere 
in a cylindrical tube.  
This model problem has been considered as 
a canonical flow in computational rheology. 
For example, a wide range of works from the
Non-Newtonian fluid mechanics community 
have investigated this flow scenario in the 
presence of shear-thinning and yield-stress 
\cite{Coussot:2014,Beris:1985,Mitsoulis:1997},
shear-thickening \cite{Tripathi:1995,Tian:2018}, 
and viscoelasticity \cite{Gervang:1992,Huang:1995}. 
When it comes to thixotropic fluids, 
both previous experimental \cite{Ferrioir:2004,Jirsaraei:2018} and 
numerical \cite{Gumulya:2014} studies have mainly focused 
on the effect of material aging time on terminal velocity (or resistance) of a spherical particle.
However, their analysis did not extend to illustrating
how fluid dynamics of a material 
will interplay with other factors of thixotropy, 
such as non-homogenous shear-breakdown 
and the convection of microstructures.
Aside from theoretical interest, 
the thixotropic flow around a sphere 
is also closely linked with various industrial 
and medical applications as well. 
Flow assurance and drilling of oil \& gas wells 
are prime examples in industry~\cite{Thant:2011}, 
and blood clotting disorder~\cite{Huang:1976}
and dysphagia (difficulty in swallowing)~\cite{Food1}
are highlighted topics in medical applications.
We believe that the current work 
will provide a useful insight for analyzing relevant flow scenarios
of ultimate interest.

The following Section~\ref{sec:formulation}
formulates a model problem 
and introduces characterizing 
dimensionless numbers.
A brief description on an employed numerical 
method is also provided.
In Section~\ref{sec:competing_factors}, 
simulation results are analyzed with 
relevant dimensionless numbers and
a general framework for 
interpreting non-homogenous 
thixotropic flow is suggested. 
With this, we investigate the effect of 
destruction factor, the viscosity ratio of 
full-structured state to broken state, and confinement 
to a non-homogenous thixotropic flow
in Section~\ref{sec:other_effect}. 
Cases of nonlinear thixotropic models are
presented in Section~\ref{sec:nonlinear}.
Lastly, Section~\ref{sec:conclusion} summarizes
the overall result and discuss future research direction
for non-homogenous thixotropic flow. 

\section{Formulation}
\label{sec:formulation}

\subsection{Structure-kinetics model}

\begin{figure}
\begin{center}
\begin{tikzpicture}
\begin{loglogaxis}
[ 
ylabel={$\eta_{\mathrm{ss}} (\dot\gamma) /(\eta_\infty+\eta_{str})$ [-]},
xlabel={$t_c \dot{\gamma}$ [-]},
xmin=2*10^(-3),
xmax=5*10^5,
width=0.6\textwidth,
]
\addplot[no marks, ultra thick, color=red, mark size =2.0]
table[x expr=\thisrowno{0},y expr=\thisrowno{1},y error expr=\thisrowno{2}] 
{Figures/vis_ss_of_model.txt};
\node at (axis cs:900, 0.5)
{\large $\eta(\dot{\gamma})=\eta_{\infty}+
\frac{\eta_{str}}{t_c \dot\gamma + 1} $};
\addplot[mark=none, black, dashed] coordinates {(10^(-4),0.00990) (10^6,0.00990)};
\node at (axis cs:0.5, 0.015)
{$\xi \left(=\frac{\eta_{\infty}} {\eta_\infty+\eta_{str}}\right)=0.00990$};
\end{loglogaxis}
\end{tikzpicture}
\end{center}
\caption{The steady state viscosity $\eta_{\mathrm{ss}}$ 
(\ref{e:eta_ss}) of Moore
thixotropy fluids in simple shear with the model parameters 
$\{\eta_\infty, \eta_{str},k_a,k_d \}$
summarized in Table~\ref{tab:model_parameter}.
The vertical line is normalized by $\eta_{\infty}+\eta_{str}$
and horizontal line is by the characteristic time of the structure formation $t_c(=k_d/k_a)$.
As the strength of shear flow increases,  $\eta_{\mathrm{ss}}$ 
converges to the viscosity ratio $\xi=0.00990$~(\ref{e:xi})
of completely-broken structure to
full structure. 
}
\label{fig:vis_ss_of_model}
\end{figure}
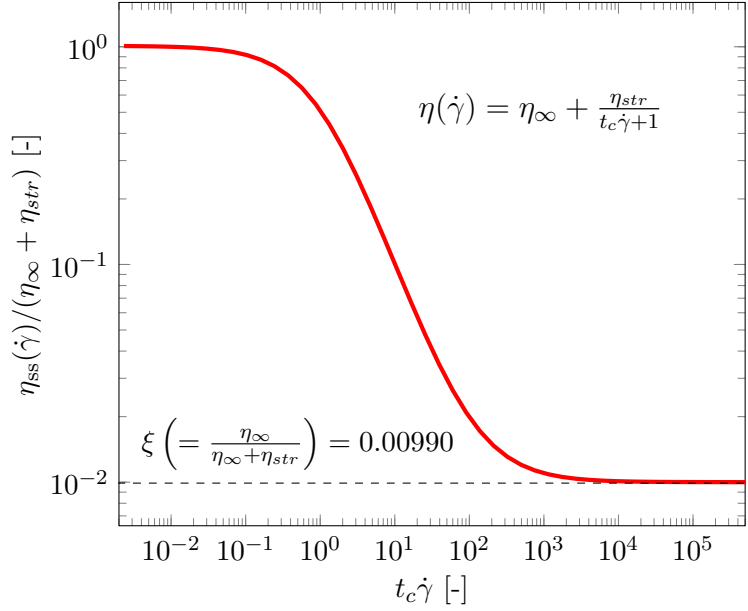

To model a generic thixotropic fluid, 
we employ the structure kinetics approach to thixotropy. 
In this approach, the instantaneous 
material structure is represented by means of
a dimensionless scalar parameter $\lambda$,
which determines the rheological properties of a material.
Then, a structural evolution is introduced 
to delineate the rate of change in $\lambda$ 
as a function of flow condition and current status of microstructure.
Although the structure kinetics approach has some 
difficulty in clearly correlating 
the structure parameter $\lambda$
to experimental measurements of the microstructure,
it is favored for a generic thixotropy model 
because of the wide range and the complex 
nature of microstructures that can possibly be 
encountered in different types of thixotropic materials \cite{Mewis:2006}.
Indeed, it has been applied to a wide range of materials 
and provided effective interpretation and prediction of thixotropy.

In general, structure-kinetics models can be classified
according to (a) the kinetic evolution equation for $\lambda$,
(b) the basic constitutive equation,
and (c) the manner in which the rheological parameters
have been linked to the structural parameter \cite{Mewis:2011}.
In this work, we take a simple model-form 
that describes the basic thixotropic features. 
To begin with, we consider the linear structure-kinetics model \cite{Moore:1959}
in a flow of velocity $\bu$, 
\begin{equation}
\frac{d\lambda}{dt}=\frac{\partial \lambda}{\partial t} + (\bu \cdot \nabla \lambda) 
=-k_d \dgamma \lambda + k_a (1-\lambda),
\label{e:Toorman_kinetic}
\end{equation}
where $\dgamma$ is the second-invariant of strain-rate tensor 
($\bgamma=\nabla \bu +(\nabla \bu)^{T}$), 
\begin{equation}
\dgamma = \sqrt{\frac{\bgamma:\bgamma}{2}}.
\end{equation}
In Eq.~(\ref{e:Toorman_kinetic}), 
$k_a$ [1/s] describes the rate of  
Brownian recovery and 
the destruction parameter $k_d$ [-] 
is related to structure sensitivity to applied shear-rate. 
In this model, $\lambda$ is restricted to $\lambda \in [0,1]$;
$\lambda=0$ denotes completely broken state
and $\lambda=1$ represents fully recovered 
microstructure respectively.  

Another equation of a structure-kinetics model 
is a constitutive equation. 
We consider a purely viscous (non-elastic)
Moore fluid \cite{Moore:1959},
in which the viscosity of a material depends on $\lambda$ as
\begin{equation}
\eta(\lambda) = \eta_\infty + \eta_{str} \lambda.
\label{e:Moore_viscosity}
\end{equation}
Here, $\eta_{str}$ expresses the structural 
contribution to the viscosity and
$\eta_\infty$ the residual viscosity 
when the microstructure is completely broken 
down ($\lambda=0$). 
In the form~(\ref{e:Moore_viscosity}), it is assumed that 
the high shear limiting behavior is expected to be Newtonian, 
ignoring inter-particle forces
once the reversible floc structures 
have been destroyed completely \cite{Mewis:2009}. 
From now on, we will simply call this combination of model 
Eq.~(\ref{e:Toorman_kinetic}) and Eq.~(\ref{e:Moore_viscosity})
as a Moore thixotropy model. 
This relatively simple model 
has been employed for computer simulation
and theoretical studies of thixotropic fluids
\cite{Gumulya:2014,Derksen:2010,Freund:2018}. 
It also has been adopted to model and predict real materials:
thixotropy in blood \cite{Moore_example1} 
and drilling fluids \cite{Moore_example2}. 

The parameters $\{k_d,k_a,\eta_\infty, \eta_{str}\}$ 
of the Moore thixotropy model,
Eq.(\ref{e:Toorman_kinetic}) and Eq.(\ref{e:Moore_viscosity}),
are usually calibrated from 
data (e.g., transient shear data) 
collected at simplest geometries from a rheometer
\cite{Mewis:2011}.
The Moore model has a characteristic time of 
structure formation $t_c=k_d/k_a$.
In a homogenous shear flow,
where no gradient of $\lambda$ exists 
in $\bu$ direction ($\bu\cdot \nabla\lambda=0$), 
the homogenous solution of material structure is 
determined by $t_c$ as
$\lambda_{ss}=1/(t_c\dot{\gamma}+1)$.
Thus, the long term behavior of the Moore model at simple shear test 
can be found by substituting 
$\lambda_{\mathrm{ss}}$ to Eq.~(\ref{e:Moore_viscosity}),
\begin{equation}
\eta_{\mathrm{ss}}(\dot\gamma)=\eta_\infty +
\frac{\eta_{str} }{t_c\dot{\gamma}+1}.
\label{e:eta_ss}
\end{equation}
Figure~\ref{fig:vis_ss_of_model} shows the steady state 
viscosity $\eta_{\mathrm{ss}}$ as a function of applied shear-rate 
$\dot{\gamma}$ in dimensionless scales.
The viscosity is normalized by $\eta_{\infty}+\eta_{str}$
and the shear rate by $t_c$.
The former is the normalized viscosity with respect to 
viscosity of full-structured material 
and the latter compares the flow strength to material time scale. 
Note that while there are infinite combinations of $k_a$ and $k_d$
to simultaneously fit steady shear data, $t_c$ will remain as a constant.

\begin{figure}
\begin{center}
\includegraphics[width=0.4\textwidth]{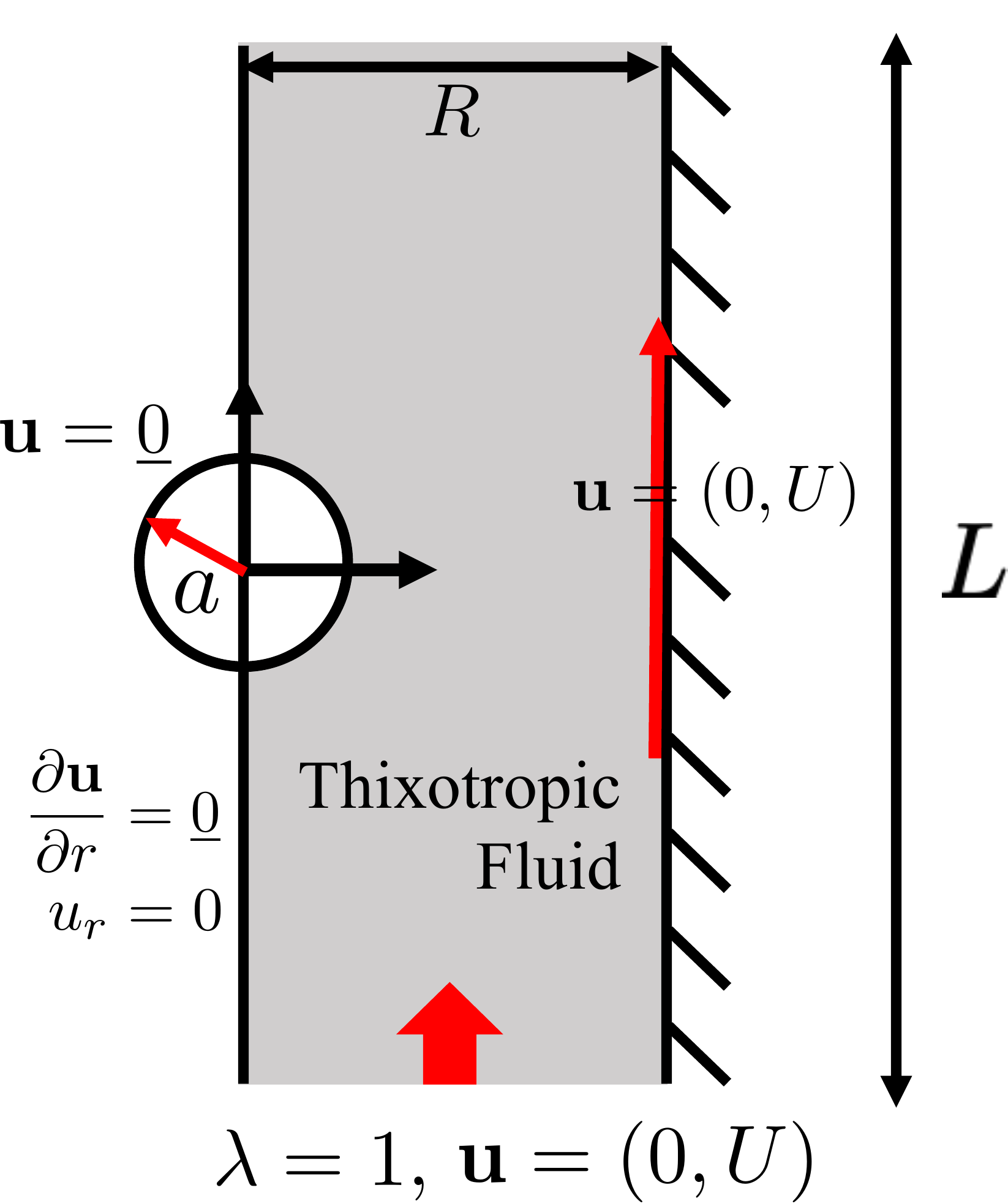}
\end{center}
\caption{Schematic representation of a settling sphere inside a 
tube filled with a fluid. 
The boundary conditions for $\mathbf{u}=(u_r,u_z)$ are described on the frame of the sphere. 
The simulation domain for the axisymmetrically constrained case is shaded.}
\label{fig:schematic}
\end{figure}

\subsection{Flow scenario}

We consider a thixotropic flow (with density $\rho_{\mathrm{f}}$)
around a solid sphere settling under gravity $g$. 
The sphere has radius $a$ and is falling at constant velocity $U$.
As shown in the schematic Figure~\ref{fig:schematic}, 
the flow is confined by an infinitely long tube (radius $R$)
filled with fully structured ($\lambda(\bx)=1$) thixotropic material. 
For many multi-particle systems (e.g., colloids and suspensions), 
viscous forces dominate the advective inertia forces, 
and thus inertia effects are 
not a practical concern~\cite{Happel:1965}.
Thus, we assume Stokes flow (or creeping flow), 
which characterizes a slow flow 
with high viscosity and small-length scale.
(i.e. $\mathrm{Re} \ll 1$). 

The incompressible 
Stokes flow around a sphere is governed by 
momentum and mass conservations,
\begin{equation}
\nabla \cdot \btau - \nabla p =0, 
\label{e:gv1}
\end{equation}
\begin{equation}
\nabla \cdot \bu =0.
\label{e:gv2}
\end{equation}
Here, $p=p^{a}+\rho_\mathrm{f} g z$  is the modified pressure, 
which is the sum of ambient pressure $p^{a}$  and
hydrostatic pressure $\rho_\mathrm{f} g z$. 
We assume the flow is axisymmetric. 
With a cylindrical coordinate, 
the flow around a 3-dimensional sphere in a cylindrical pipe
reduces to 2-dimensional flow $(u_r,u_z)$, because $u_\theta=0$
and $\partial/\partial \theta=0$.

The steady state ($\partial/\partial t =0$) structure-kinetics equation 
\begin{equation}
(\bu \cdot \nabla \lambda)+k_d \dgamma \lambda - k_a (1-\lambda)=0,
\label{e:gv3}
\end{equation}
is coupled with the momentum conservation,
in a way that the material stress $\btau$ is described by 
the constitutive equation with Moore viscosity function 
$(\ref{e:Moore_viscosity})$
\begin{equation}
\btau=(\eta_{\infty}+\eta_{str} \lambda) \bgamma.
\label{e:constitutive_eq}
\end{equation}

The boundary conditions are 
described in the frame fixed at the center of the sphere,
\begin{equation}
u_z=U \quad \textrm{at}\quad r=R\; \textrm{ or }  z=\pm \frac{1}{2}L, 
\end{equation}
\begin{equation}
u_r=u_z=0 \quad \textrm{at}\quad \sqrt{r^2+z^2}=a.
\end{equation}
with axisymmetric constraints, 
\begin{equation}
u_r=0 \textrm{ and } \frac{\partial u_r}{\partial r}=0 \quad \textrm{at}\quad r=0.
\end{equation}

In the current frame, 
the fully structured ($\lambda=1$) material 
is being convected with velocity $U$ 
from the bottom boundary
(in-flow boundary).
Thus, the boundary condition for Eq.~(\ref{e:gv3}) is
\begin{equation}
\lambda = 1 \quad \textrm{at}\quad z=-\frac{1}{2}L. 
\end{equation}
The aforementioned boundary conditions are summarized in 
the schematic Figure~\ref{fig:schematic}.

In this flow configuration, we define a quantity of interest
as the resistance $\cD$ of the falling sphere, 
which can be calculated by
the surface integral over the sphere surface $\partial S$,

\begin{equation}
\cD = \int_\pS \mathbf{e}_z \cdot (-p \bI + \btau) \cdot \bn \, dA,
\label{e:cD}
\end{equation}
where $\mathbf{e}_z$ is the unit vector in $z$-direction, 
$\bn$ is the normal vector, 
and $\bI$ is the identity tensor.

\subsection{Parameters and Dimensionless numbers}
\begin{table}
\begin{center}
\begin{tabular}{|c|c|}
\hline
$\eta_{str}$      & 100.0 Pa$\cdot$s \\
\hline
$\eta_{\infty}$  & 1.0 Pa$\cdot$s \\
\hline
$k_a$           & $0.1\;\mathrm{s}^{-1}$ \\
\hline
$k_d$     & $1.0$ \\
\hline
$t_c$      & 10 s $(=k_d/k_a)$\\
\hline
\end{tabular}
\caption{Thixotropic model parameters used
in simulation}\label{tab:model_parameter}
\end{center}
\end{table}

The flow problem is solved by a numerical solver,
which will be introduced later in Sec~\ref{sec_Nscheme}.
The numerical scheme has a set of input parameters 
$\{\eta_\infty,\eta_{str},k_d,k_a ; a, R, L ,U\}$.
Here, the first four parameters are the rheological model parameters
and represent the intrinsic property of a material.
The other four are flow-scenario parameters
and denote the boundary condition or geometry. 
Model parameters and scenario parameters are independent; 
one may consider various flow scenarios for a single material 
at any time. 
Since the main goal of this paper 
is to illustrate how the material intrinsic thixotropy
interplays with a non-homogenous flow, 
we first fix the model parameters as
described in Table~\ref{tab:model_parameter}
in the subsequent analysis,
unless otherwise stated. 

The confining cylinder has radius $R=0.1$ m.
Ideally, we consider a sphere settling in infinitely long tube. 
However, it is impractical for the domain exterior 
$L$ to extend out to infinity in simulations. 
Therefore, the computational domain is cut off by large but finite length $L_0$. 
The simulation domain has $L_0/R=8$. 
It is confirmed that our numerical results do not change 
with the variation of this parameter.
At first, the sphere radius is set as $a=0.025$ m, 
and in Sec~\ref{sec_confinement}, the effect of confinement~$(a/R)$ 
will also be discussed separately.

Now, the model thixotropic fluid flow 
can be characterized by a few dimensionless numbers. 
First, the viscosity ratio $\xi$ of 
completely-broken structure to
full structure is
\begin{equation}
\xi =\frac{\eta_{\infty}}{\eta_\infty+\eta_{str}}=0.00990.
\label{e:xi}
\end{equation}
We normalize the velocity of a falling sphere using 
$t_c$  and the length scale $a$ as
\begin{equation}
U^{*}=\frac{t_c U}{ a}
\label{e:dimensionless_U}
\end{equation} 
$U^{*}$ indicates external flow-strength compared to material timescale.
In numerical simulations, 
we will investigate the effect of thixotropy at different time scale by changing $U$.
This is equivalent to consider a range of different mass of sphere,
because at steady state, a force balance yields 
\begin{equation}
\cD (U) = \frac{4}{3}\pi a^3 (\rho_s-\rho_f)g,
\end{equation}
where $\rho_s$ is the density of a settling sphere. 
In experiment, it is often achieved through a ``force-control''
perspective, i.e. changing the density of a sphere
and measuring the terminal velocity \cite{Bruyn:2007}.
We also define the drag coefficient Cs of a thixotropic fluid as
\begin{equation}
\mathrm{Cs} \equiv \frac{\cD}{K6\pi (\eta_\infty+\eta_{str}) a U},
\label{e:Cs}
\end{equation}
where $\cD$ is the resistance acting on the sphere
and $K=K(a/R)$ is the wall correction factor given as 
a Faxen series for $a/R$ \cite{Happel:1965}
\begin{equation}
\begin{aligned}
K=&\bigg[ 1.0 - 2.10444\left(\frac{a}{R}\right)
+2.0877\left(\frac{a}{R}\right)^{3}
-0.94813\left(\frac{a}{R}\right)^{5}\\
&-1.372\left(\frac{a}{R}\right)^{6}
+3.87\left(\frac{a}{R}\right)^{8}
+3.87\left(\frac{a}{R}\right)^{10}
\bigg]^{-1}.
\end{aligned}
\label{e:wall_correction}
\end{equation}
The inclusion of $K$ in the definition of Cs 
is useful to compare $\cD$ in thixotropy fluid
to the existing Newtonian analytic solution,
and hence to isolate 
the effect of thixotropy on resistance. 
For example, 
numerically calculated $\cD$ through the form~(\ref{e:cD})
can be rewritten as $\cD=K6\pi \eta^{e}U a$ 
by introducing an effective viscosity 
$\eta^{e}=\cD/(K6\pi U a)$.
In this case, $\eta^{e}$ represents the gross viscosity 
acting on the sphere from the flow field and 
carries information on structural profile $\lambda(\bx)$
associated thixotropic behavior. 
Then, the form of Cs~(\ref{e:Cs}) is equivalent to
the ratio of the effective viscosity $\eta^{e}$ 
to the viscosity of fully-structured state $\eta_{\infty}+\eta_{str}$. 
Thus, Cs can be understood as a measure of 
viscosity relative to that of fully structured state. 
In addition, considering that
the linear relation between $\eta$ and $\lambda$
in Eq.~(\ref{e:Moore_viscosity}),
Cs can also be understood to present the relative amount 
of preserved structure compared to full structure. 
In this work, the possible-minimum value of Cs is given as
\begin{equation}
\mathrm{Cs}_{\mathrm{min}}=\frac{\eta_\infty}{\eta_\infty+\eta_{str}}=\xi, 
\end{equation}
for the case where the sphere travels through 
the completely-broken structure.  

Lastly, we consider the effect of dimensionless 
model parameter $k_d$ and $\xi$ separately
in Section~\ref{sub_sec:destruction} and \ref{sub_sec:xi}.
At given $U^{*}$, $k_d$ determines the relative strength of 
shear-induced structure breakdown compared to structure convection effect. 
The list of dimensionless number is summarized in Table~\ref{tab:dimensionless_number}.
It is useful to note that in the limit of either $\xi \to 1$, $U^{*} \to 0$, or $k_d \to 0$, 
the resistance (and flow solution) 
converges to that of a Newtonian fluid with 
the full-structured viscosity $\eta_\infty+\eta_{str}$ and thus $\mathrm{Cs} \to 1$. 
In the other limit, $U^{*} \to \infty$ and $k_d \to \infty$,
Cs will converge to $\mathrm{Cs}_{\mathrm{min}}$.

\begingroup 
\renewcommand{\arraystretch}{1.7} 
\begin{table}
\begin{center}
\begin{tabular}{|c|c|c|}
\hline
$\xi$    &$\frac{\eta_{\infty}}{\eta_{\infty}+\eta_{str}}$  & The viscosity ratio of completely-broken structure to full structure\\
\hline
$U^{*}$ &$\frac{t_cU}{a}$  & Normalized falling velocity \\
\hline
Cs &$ \frac{\cD}{K6\pi(\eta_{\infty}+\eta_{str})aU}$ &Drag coefficient \\
\hline
$k_d$   & -  & Destruction parmeter \\
\hline
$a/R $ & - &Confinement Ratio \\
\hline
\end{tabular}
\caption{Dimensionless numbers characterizing 
the model thixotropic fluid flow.}
\label{tab:dimensionless_number}
\end{center}
\end{table}
\endgroup

\subsection{Numerical Method}
\label{sec_Nscheme}

The governing equations (\ref{e:gv1}) to (\ref{e:gv3}) 
are solved numerically 
with finite-element method (FEM)
using a C++ FEM software library 
\texttt{deal.II} \cite{Arndt:2017,Bangerth:2007}.
The primary variables are the two velocity components 
$(u_r,u_z)$, pressure $p$, 
and structure parameter $\lambda$.

We linearize the set of governing equations 
by decoupling the structure kinetics equation (\ref{e:gv3})
from the conservation equations.
We take Picard iterations form (also known as fixed-point iteration), 
which is a simple and frequently used method for nonlinear problems. 
In this method, we computes the viscosity as a function of 
the previous structure parameter $\lambda$ 
and solve for a new velocity and pressure field repeatedly.
It is known that the choice of an initial solution is
important to obtain a converging solution.
We start from a Newtonian field with 
viscosity $\eta(\bx)=\eta_\infty+\eta_{str}$
(i.e., $\lambda(\bx)=1$) and 
Picard iterations proceed: for each $k$ step, the 
incompressible flow $(\bu^{k+1},p^{k+1})$ is solved 
with $\lambda^k$ from the previous iteration 
\begin{align}
-\nabla p^{k+1} + \nabla\cdot\left[ \eta(\lambda^k)\, \bu^{k+1}\right] &= 0, \\
\nabla\cdot\bu^{k+1} &=0.
\end{align}
These equations represent a symmetric saddle point problem, 
and the function spaces should satisfy 
the Ladyzhenskaya-Babuska-Brezzi (LBB) condition. 
To satisfy this, we employ  the second-order 
Taylor--Hood elements \cite{TaylorHood}, 
composed of continuous piecewise quadratic
element for $\bu$ and continuous piecewise linear for $p$.
Schur complement approach is used for  
solving the linearized system for $(\mathbf{u},p)$.

Once the flow solution $(\bu^{k+1},p^{k})$ is found, 
the scalar $\lambda^{k+1}$ is updated according to
\begin{equation}
\bu^{k+1} \cdot\nabla \lambda^{k+1} 
+ k_d \dgamma\,\lambda^{k+1} - k_a \,(1-\lambda^{k+1})=0.
\label{e:gv3_linearized}
\end{equation}
As standard Galerkin method is numerically unstable 
for this advection type equation~(\ref{e:gv3_linearized})
discontinuous Galerkin (DG) approximation
with degree 2 is used for $\lambda(\bx)$ 
\cite{DGbook,Saramito:2016_book}. 
The advantages of using DG method is that 
it solves hyperbolic system using FEM 
without artificial viscosity for stabilization.
In complex fluid flows simulation, 
the method has been often adopted for viscoelastic flows 
which has stress being convected in the domain
\cite{Saramito:2016_book}.  

The weak form for Eq.~(\ref{e:gv3_linearized})
is defined locally at each element $K$ 
in domain $\Omega$. 
To derive it,
we multiply the test function $w$
on both sides of Eq.~(\ref{e:gv3_linearized})
and take integral by part on the advection term, 

\begin{equation}
\int_{K} \big[ w (k_d \dgamma+k_a)\lambda^{k+1} 
+(\bu^{k+1} \cdot \nabla w)\lambda^{k+1}- k_a \, w \big] dV
+ \int_{\partial K} \big[\bu\cdot (w^{+}\bn^{+}+ w^{-}\bn^{-})\lambda^{k+1} \big] dS
=0.
\label{e:dgweak}
\end{equation}
Here, the superscripts denote 
the upwind (-) and downwind (+) 
values at the internal faces respectively.
For numerical stability, 
we select upwind flux $\big(\lambda^{k+1}\big)^{-}$
for face integrals. 
The boundary condition ($\lambda=1$) at the inlet 
replace the second integral term 
for faces located at $z=-L/2$.
The bilinear form for Eq.~(\ref{e:dgweak}) 
is solved by GMRES solver \cite{GMRES}
with a preconditionner constructed by incomplete LU
factorization \cite{ILU}. 

A single simulation runs with
49,152-elements (mesh) resulting the number of degree of freedom 
for 395,010 for $\bu$, 49,601 for $p$, and 196,608 for $\lambda$ in total.
The convergence criteria $\zeta=10^{-8}$ used for Picard iteration is 
relative $L_2$-norm difference of $(u_r,u_z,p,\lambda)$
between subsequent $k$ iteration.  
It is confirmed that our results do not change with the variation of 
$\zeta$.
To verify the present numerical scheme, 
we used the method of manufactured solution, 
whose procedure and result are summarized in Appendix~\ref{a:code_verification}.

\section{Thixotropy determined by interplay of three competing factors}
\label{sec:competing_factors}

\begin{figure}
\begin{center}
\begin{tikzpicture}
\draw (0, 0) node[inner sep=0]{
\includegraphics[width=1.0\textwidth,trim=50 322 20 300 mm, clip=true]
{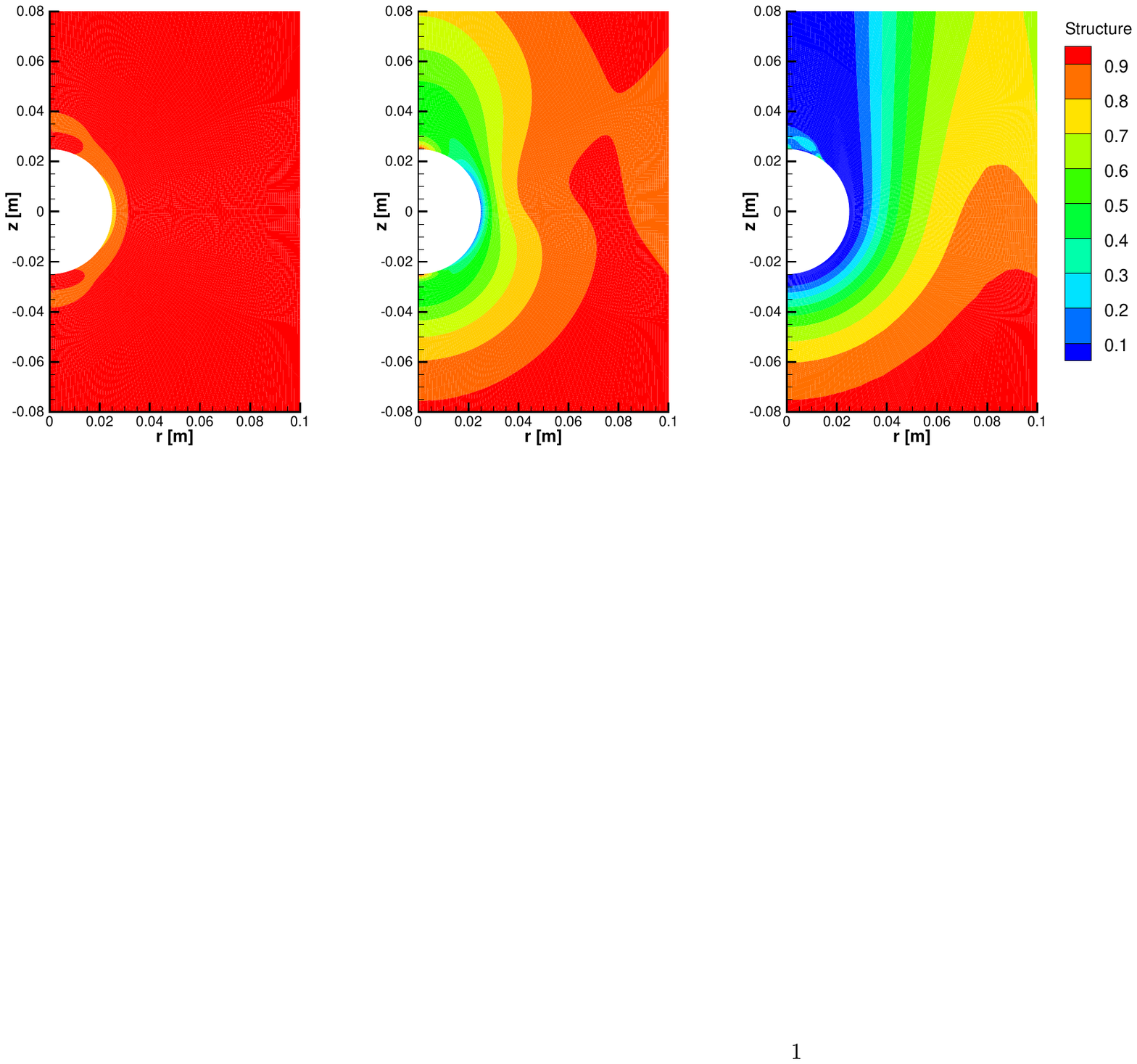}}; 
\draw (-5.60, 3.1) node {(\textbf{a}) $U^{*}=0.1$};  
\draw (-0.5, 3.1) node {(\textbf{b}) $U^{*}=1.0$};  
\draw (4.7, 3.1) node {(\textbf{c}) $U^{*}=100$};  
\end{tikzpicture}
\end{center}
\caption{Structure $\lambda(\bx)$ at $U^{*}=0.1,1.0,100$
at confinement ratio $a/R=0.25$
$U^{*}=(t_c/a)U$ represents the ratio of thixotropic time
scale to the flow strength.
When $U^{*}=0.1$, almost full structure is maintained and solution 
is close to Newtonian with viscosity $\eta_\infty+\eta_\mathrm{str}$.
As $U^{*}$ increases, broken structures (blue region) start to appear
from the vicinity of the sphere $(r,z)=(a,0)$.}
\label{fig:lambda_for_u}
\end{figure}

Thixotropic behavior of a Moore fluid flow 
around the solid sphere is governed by 
the structural kinetic evolution of $\lambda(\bx)$. 
At steady state, 
$\lambda(\bx)$ solution is determined by 
Eq.~(\ref{e:gv3}), which consists of three factors:
structure recovery by Brownian motion,
shear-induced breakdown, and the convection of microstructure.
Each factor has its own time scale. 
Firstly, the model parameter $k_a$ [$\mathrm{s}^{-1}$] represents 
the time scale for Brownian structure recovery, 
which has nothing to do with the flow solution $\bu$.
Secondly, the time scale for shear-induced breakdown
varies at each space due to its dependency on $\bu$. 
Considering the characteristic shear-rate $U/a$ of the flow, 
the breakdown timescale for overall domain is
estimated as 
$(Uk_d)/a$ [$\mathrm{s}^{-1}$]. 
It is worth noting that the normalized velocity (\ref{e:dimensionless_U})
is the ratio of the time scale for shear-induced breakdown 
to that of structure recovery. 
Lastly, the time scale for 
the convection is related to $U/a$ [$\mathrm{s}^{-1}$]
because the flow rate of convecting material is proportional to 
$U$ and a sphere with smaller radius $a$ is more easily enveloped
by the same amount of convecting material. 
The time scale of both structure-convection and shear induced breakdown 
decreases in the same scale, when $U$ is increased. 

\begin{figure}[t]
\begin{center}
\begin{tikzpicture}
\draw (0, 0) node[inner sep=0]{
\includegraphics[width=1.0\textwidth,trim=60 335 60 320 mm, clip=true]
{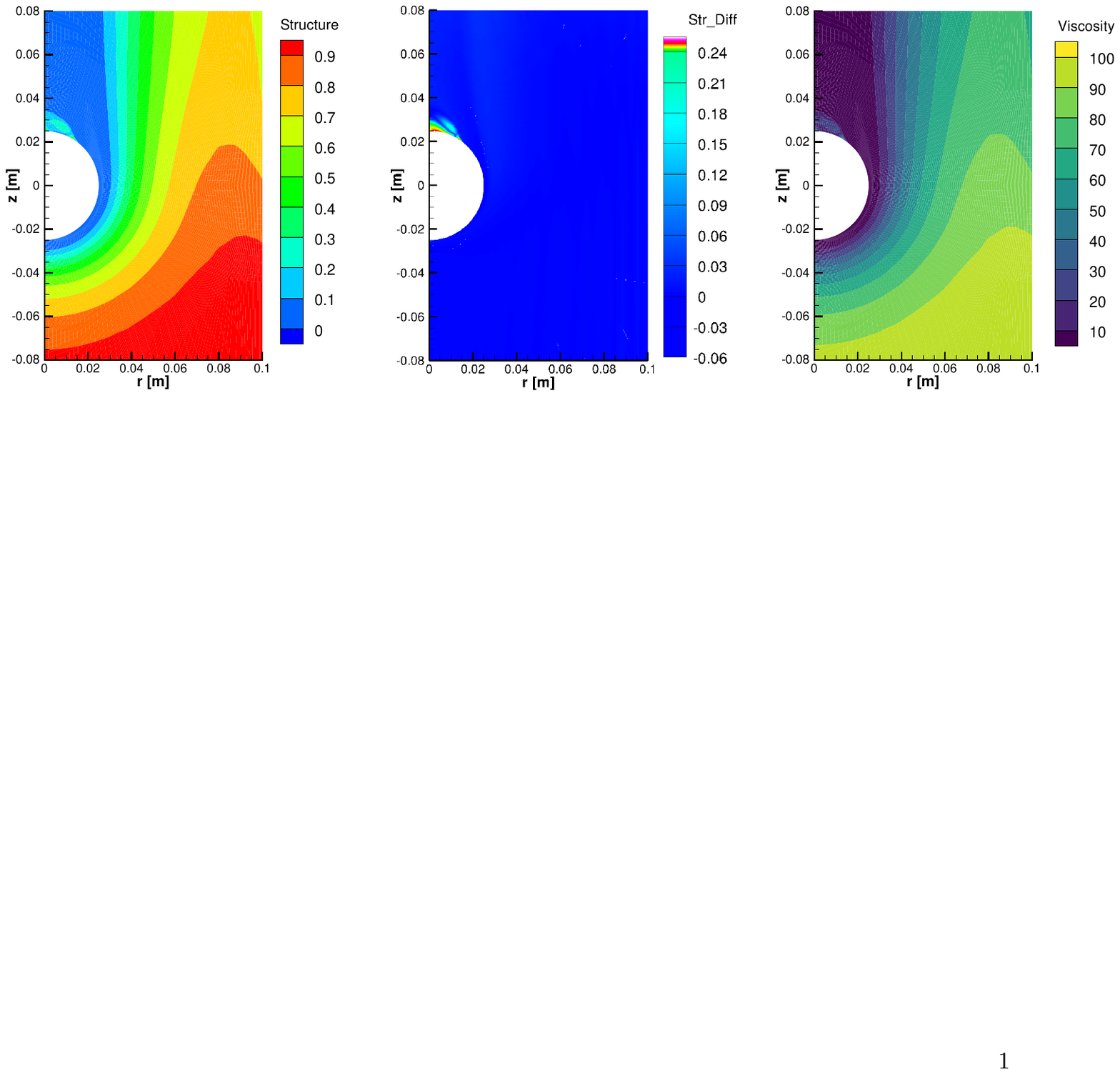}}; 
\draw (-5.8,2.8) node {(\textbf{a}) $\lambda_{U^{*}=200}(\bx)$};  
\draw (-0.3, 2.8) node {(\textbf{b}) $\Delta \lambda(\bx)$};  
\draw (5.3, 2.8) node {(\textbf{c}) $\eta_{U^{*}=200}(\bx)$ [Pa$\cdot$s]};  
\end{tikzpicture}
\end{center}
\caption{(\textbf{a}) $\lambda(\bx)$ at $U^{*}=200$.
(\textbf{b}) Difference in structure solution 
between $U^{*}=200$ and $U^{*}=100$.
Due to structure convection, 
the solution shape of $\lambda$ 
is maintained qualitatively similar to that of 
$U^{*}=100$ in Figure~\ref{fig:lambda_for_u}\textbf{c}.
(\textbf{c}) Corresponding viscosity field $\eta(\bx)=\eta_\infty+\eta_{str} \lambda(\bx).$ 
}
\label{fig:lambda_for_u200}
\end{figure}

Figure~\ref{fig:lambda_for_u} shows the structure profile at 
three different $U^{*}$~(0.1, 1 and 100). 
When $U^{*}=0.1$, the Brownian structure recovery,
which has the shortest time scale,
is dominant over the other two factors. 
Thus, $\lambda(\bx)$ remains homogenous in the flow field 
with almost full structure $\lambda \approx 1$. 
For example, $\lambda_{U^{*}=0.1}(\bx)$ shown in Figure~\ref{fig:lambda_for_u}(a)
has just a little inhomogeneity just around the sphere. 
As $U^{*}$ increases, the other two factors start to play an important
role in determining $\lambda(\bx)$. 
The strengthened shear-induced structure breakdown 
effect is manifested by non-homogenous $\lambda_{U^{*}=1}(\bx)$
profile as shown in Figure~\ref{fig:lambda_for_u}(b). 
As $U^{*}$ increases further, the effect of Brownian structure recovery 
becomes negligible compared to other two factors.
When $U^{*}\gg1$,
the shear-induced breakdown effect and
the convection of full microstructure 
reach an equilibrium,  
since both factors increase with the same scale to $U^{*}$.
Consequently, a further increase in $U^{*}$ does not cause 
remarkable change in $\lambda(\bx)$ profile. 
For instance, $\lambda(\bx)$ at $U^{*}=200$ in Figure~\ref{fig:lambda_for_u200}(a)
and $U^{*}=100$ in Figure~\ref{fig:lambda_for_u}(c)
demonstrate barely any difference as displayed in Figure~\ref{fig:lambda_for_u200}(b). 
Here, the minute difference at the wake of the sphere
is attributed to the fact that fluid elements in this region are mostly subjected 
to extensional stress rather than shear stress; 
even a higher velocity $U$ cannot lead enough shear
to induce completely breakdown structure behind the sphere. 
Also, the fluid elements in this region
are not replaced by the convecting structured material
from the front of the sphere due to the no-slip condition
on the sphere surface. 
Therefore, a Brownian recovery effect
still plays a significant role even at larger $U^{*}$.
The corresponding viscosity field reconstructed 
from $\lambda(\bx)$ is shown in Figure~\ref{fig:lambda_for_u200}(c).

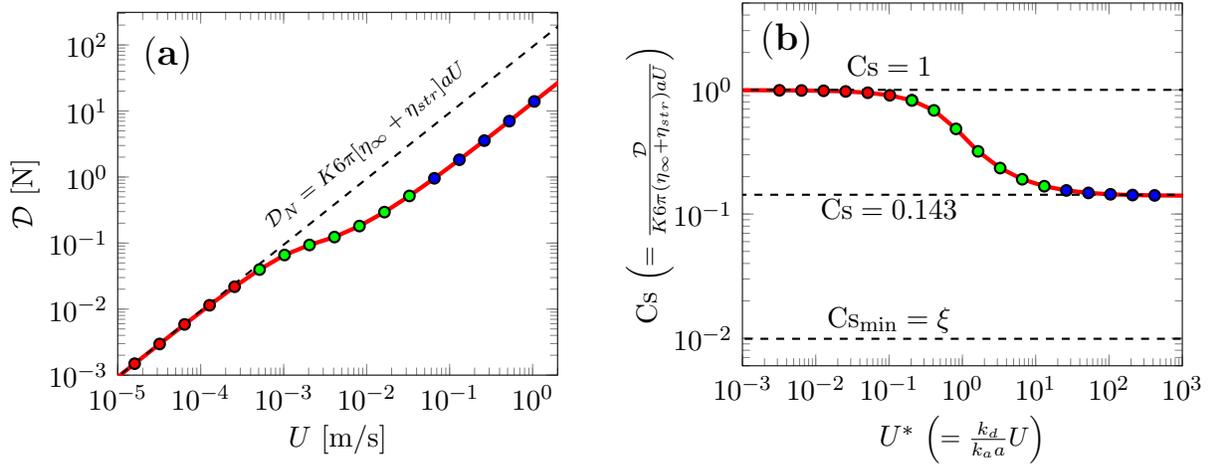
\begin{figure}
\begin{subfigure}{0.45\textwidth} 
\begin{center}
\begin{tikzpicture}
\begin{loglogaxis}
[ 
ylabel={$\cD$ [N]},
xlabel={$U$ [m/s]},
xmin=1*10^(-5),
xmax=2,
ymin=10^(-3), 
width=1.0\textwidth,
]
\node at (axis cs:0.00004, 70) {\Large (\textbf{a})};
\addplot[no marks, ultra thick, color=red, mark size =2.0]
table[x expr=\thisrowno{0},y expr=\thisrowno{1},y error expr=\thisrowno{2}] 
{Figures/D_U_a025.txt};
\addplot[mark=*, mark options ={fill=red}, only marks, mark size =2,  thick, color=black]
 coordinates { 
(8.0e-06	, 0.00074701)
(1.6e-05,  0.00148957)
(3.2e-05,  0.0029614)
(6.4e-05,  0.00585246)
(0.000128,  0.0114275)
(0.000256,  0.0217987)
};
\addplot[mark=*, mark options ={fill=green}, only marks, mark size =2,  thick, color=black]
 coordinates { 
(0.000512,  0.0396895)
(0.001024,  0.0659128)
(0.002048,  0.0935708)
(0.004096,  0.123334)
(0.008192,  0.18094)
(0.016384,  0.294195)
(0.032768,  0.517115) 
};
\addplot[mark=*, mark options ={fill=blue}, only marks, mark size =2,  thick, color=black]
 coordinates { 
(0.065536,  0.956215)
(0.131072,  1.82659) 
(0.262144,  3.55982) 
(0.524288,  7.02402)
(1.04858000000000,	13.9542000000000)
};
\addplot[dashed, thick, color=black]
table[x expr=\thisrowno{0},y expr=\thisrowno{1},y error expr=\thisrowno{2}] 
{Figures/Newton_D_U_a025.txt};
\node[rotate=39] at (axis cs:0.01, 3)
{\scriptsize $\cD_N=K 6 \pi  [\eta_\infty + \eta_{str}] a U$};
\end{loglogaxis}
\end{tikzpicture}
\end{center}
\end{subfigure} \hspace{0.03\textwidth}
\begin{subfigure}{0.45\textwidth} 
\begin{center}
\begin{tikzpicture}
\begin{loglogaxis}
[ 
ylabel={Cs \small $\left(=\frac{\cD}{K6\pi (\eta_\infty+\eta_{str}) a U} \right)$},
xlabel={$U^{*}$ \small $\left(=\frac{k_d }{k_a a} U \right)$},
xmin=10^(-3),
xmax=1000,
ymin=0.006, 
ymax=5,
width=1.0\textwidth,
]
\node at (axis cs:0.004, 2.5) {\Large (\textbf{b})};
\addplot[no marks, ultra thick, color=red, mark size =2.0]
table[x expr=\thisrowno{0},y expr=\thisrowno{1},y error expr=\thisrowno{2}] 
{Figures/Cs_Wi_a025_revised.txt};
\addplot[dashed, thick, color=black]
table[x expr=\thisrowno{0},y expr=\thisrowno{1},y error expr=\thisrowno{2}] 
{Figures/Cs_Wi_a025_Newton.txt};
\node at (axis cs:0.1, 1.55)
{$\mathrm{Cs}=1$};
\addplot[dashed, thick, color=black]
table[x expr=\thisrowno{0},y expr=\thisrowno{1},y error expr=\thisrowno{2}] 
{Figures/Cs_Wi_a025_Newton2.txt};
\node at (axis cs:0.1, 0.11)
{$\mathrm{Cs}=0.143$};
\addplot[dashed, thick, color=black]
table[x expr=\thisrowno{0},y expr=\thisrowno{1},y error expr=\thisrowno{2}] 
{Figures/Cs_Wi_a025_Newton3.txt};
\node at (axis cs:0.1, 0.014)
{$\mathrm{Cs}_\mathrm{min}=\xi$};
\addplot[mark=*, mark options ={fill=red}, only marks, mark size =2,  thick, color=black]
 coordinates { 
(3.20000000000000e-03,	0.991402648094799)
(6.40000000000000e-03,	0.988449714543694)
(0.0128000000000000,	0.982563754858682)
(0.0256000000000000,	0.970894690477518)
(0.0512000000000000,	0.947883375147531)
(0.102400000000000,	0.904074615175169)
(0.204800000000000,	0.823036911352394)
};
\addplot[mark=*, mark options ={fill=green}, only marks, mark size =2,  thick, color=black]
 coordinates { 
(0.204800000000000,	0.823036911352394)
(0.409600000000000,	0.683413337665983)
(0.819200000000000,	0.485091914894195)
(1.63840000000000,	0.319695493848298)
(3.27680000000000,	0.234508337753219)
(6.55360000000000,	0.190646569098343)
(13.1072000000000,	0.167552474683950)
};
\addplot[mark=*, mark options ={fill=blue}, only marks, mark size =2,  thick, color=black]
coordinates { 
(26.2144000000000,	0.154913500459195)
(52.4288000000000,	0.147960161053613)
(104.857600000000,	0.144178918236132)
(209.715200000000,	0.142242529856700)
(419.432000000000,	0.141291819957534)
};
\end{loglogaxis}
\end{tikzpicture}
\end{center}
\end{subfigure}
\caption{(\textbf{a}) Prediction of $\cD$ for various $U$. 
The dashed line is Newtonian resistance $\cD_N$
with viscosity $\eta_\infty+\eta_{str}$.
(\textbf{b}) The relation between Cs and $U^{*}$. 
The vertical line represents 
gross viscosity (normalized by $\eta_\infty+\eta_0$)
that the sphere experiences by the viscosity field $\eta(\bx)$. 
The shape of the Cs-$U^{*}$ curve is qualitatively similar to the flow curve 
(Figure~\ref{fig:vis_ss_of_model}).
The Cs value at the terminal regime (blue points) 
does not equal to $\xi$ due to the structure convection. 
}
\label{fig:cs_to_ustar}
\end{figure}

For more quantitative analysis on the three competing 
factors, the resistance of the sphere is investigated. 
The distribution profile of $\lambda(\bx)$ in Figure~\ref{fig:lambda_for_u}
takes a 2-dimensional form that prevents a more quantitative 
description and intuitive understanding of thixotropy 
in the current problem. 
Recognizing that the resistance $\cD$, 
which can be calculated through the form~(\ref{e:cD}),
implicitly carries the information on $\lambda(\bx)$, 
we interpret that 
$\cD$ represents a measure of the distribution of $\lambda(\bx)$. 
The transition between three competing factors will
therefore also be reflected in this single parameter. 
Figure~\ref{fig:cs_to_ustar}(a) shows $\cD$
as a function of settling speed $U$. 
The scale of Reynolds number (Re) in Figure~\ref{fig:cs_to_ustar}(a) varies,
since we considered a wide range of settling velocity $U$. 
While Re is fully determined if the fluid density $\rho_{\mathrm{f}}$ is specified, 
our parametric study with respect to $U$ actually implies that 
the density difference between the sphere and fluid is varied. 
For a rough estimations on Re(=$\rho_{\mathrm{f}}Ua/\eta$), 
we first assume that $\rho_{\mathrm{f}} \approx O(1000)\; \mathrm{kg/m^3}$. 
Considering the sphere radius $a=0.025\; \mathrm{m}$ and
thixotropic viscosity 100-15$\; \mathrm{Pa\cdot s}$,  
we estimate the range of Re as $10^{-5}\; \mathrm{to} \; 0.8$.
For a higher Re regime, flow solution may be significantly 
influenced by nonlinearity coming from inertial phases.

\begin{figure}[b]
\begin{center}
\begin{tikzpicture}
\draw (0, 0) node[inner sep=0]{
\includegraphics[width=1.0\textwidth,trim=60 337 50 310 mm, clip=true]
{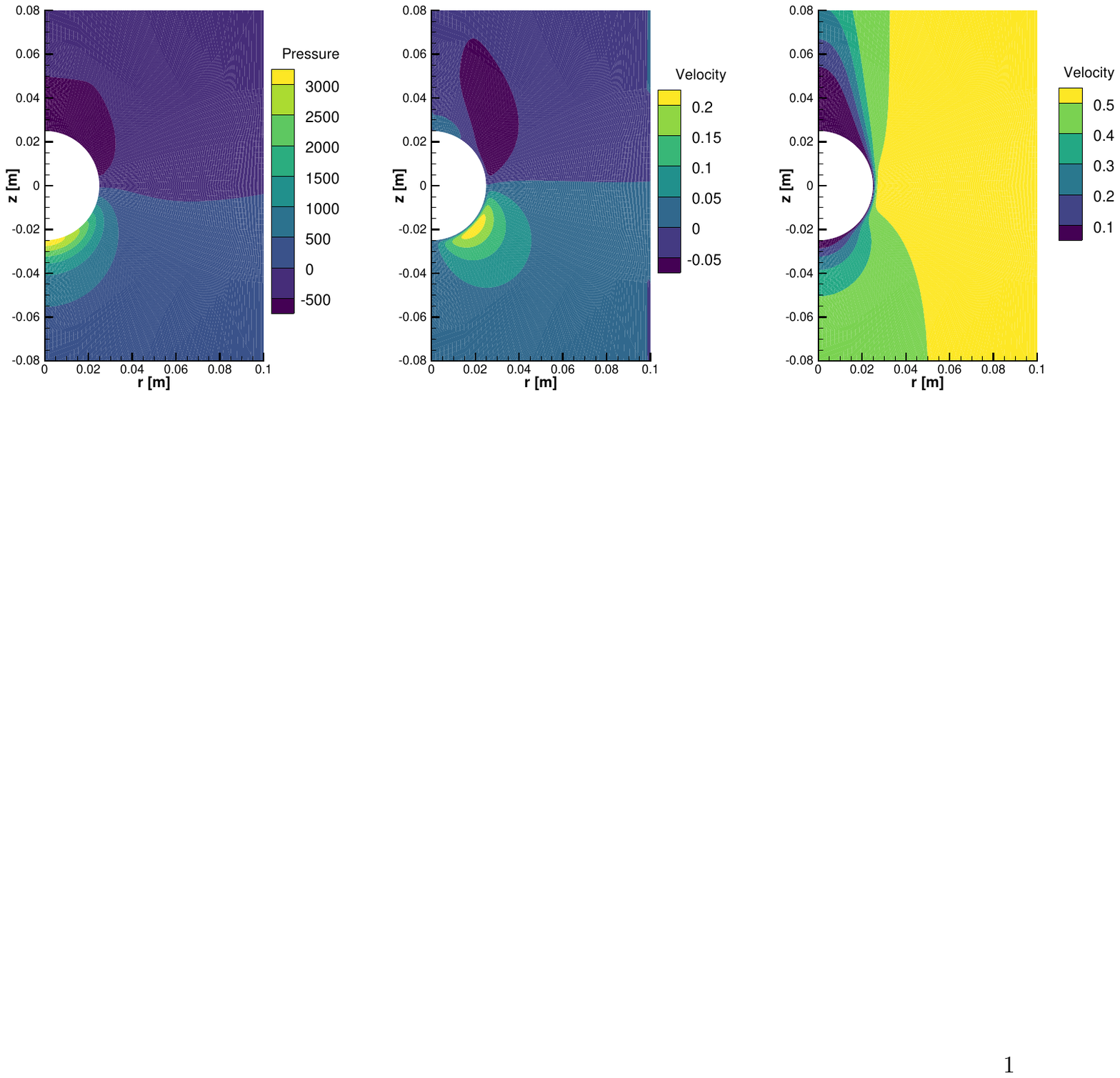}}; 
\draw (-6.0, 2.55) node {(\textbf{a}) $p$ [Pa]};  
\draw (-0.5, 2.55) node {(\textbf{b}) $u_r$ [m/s]};  
\draw (5.0, 2.55) node {(\textbf{c}) $u_z$ [m/s]};
\end{tikzpicture}
\end{center}
\caption{Flow solution $(p,u_r,u_z)$ at $U^{*}=200$.
The flow solution has fore-aft asymmetric due to 
inhomogeneous viscosity field around the sphere. 
}
\label{fig:flow_solution_for_ustar200}
\end{figure}

We also plot the relationship between 
the resistance coefficient Cs and $U^{*}$ 
in Figure~\ref{fig:cs_to_ustar}(b),
which has a form of sigmoid shape with two asymptotes. 
The curves in Figure~\ref{fig:cs_to_ustar} are divided into three regime: 
Newtonian-fluid regime at $U^{*}\ll1$,
terminal regime at $U^{*}\gg1$, 
and the transient regime in between. 
In the first regime marked with red circles, 
the thixotropic fluid behaves like a Newtonian fluid with viscosity
$\eta_\infty+\eta_{str}$ as prescribed. 
In this regime, 
$\cD$ follows the analytic resistance solution of Newtonian, 
i.e., $\cD=K6\pi (\eta_\infty+\eta_{str}) U a$. 
Likewise, the Cs in Figure~\ref{fig:cs_to_ustar}(b)
is maintained almost constant at 1, 
which implies that the thixotropic fluid is fully structured. 
As $U^{*}$ increases, 
$\cD$ shows a sub-linear increase 
(i.e. $\cD \propto U^{n}, \;\mathrm{where}\;0<n<1$), which is manifested 
by green circles in the middle of Figure~\ref{fig:cs_to_ustar}(a).  
Accordingly, Cs starts to decreases and enters 
the transient regime, which is also marked 
with the same green circles in Figure~\ref{fig:cs_to_ustar}(b).
In terms of the explanation via the three competing factors, 
the transition of Cs value is the result of a shift in the equilibrium 
of competing factors to more broken structure.
With a further increase in $U^{*}$,
thixotropic fluid finally reaches the terminal regime, 
where a balance between the structure convection
effect and the shear-induced breakdown is accomplished. 
As marked with blue circles,
$\cD$ increases linearly again with respect to $U$ 
in this regime. 
However, it should be noted that the flow is 
clearly distinguished from that of Newtonian flow,
even though $\cD$ increases linearly with $U$. 
For example, the flow solution shown at $U^{*}=200$
in Figure~\ref{fig:flow_solution_for_ustar200} has fore-aft asymmetry. 
This cannot be observed in Newtonian Stokes flow.  
This asymmetry is attributed to the effect of thixotropy. 
In the front of the sphere, the decreasing fluid viscosity 
induces a larger shear-rate and hence a greater pressure gradient. 
However, the fluid viscosity remains as a constant value $\eta_{str}$
in the wake for the sphere as shown in Figure~\ref{fig:lambda_for_u200}(c).  
This results in a relatively small gradient of pressure and velocity. 
The stronger positive pressure in the front of the sphere contributes to larger 
pressure drag. It is calculated that the pressure drag is nearly twice the viscous drag. 
It is contrasts to Newtonian drag, the viscous part of which is twice the pressure part. 
It is also observed that $u_z$ remains positive (upward) throughout the flow domain, 
indicating that there exists no negative-wake behind the sphere. 
It is conjectured that the fluid should be viscoelastic 
to generate negative wake \cite{Bird:1987,Tsamopoulos:2016}.

The origin of terminal regime is the 
result of dynamic equilibrium of 
shear-induced breakdown and the convection of structured fluid,
which does not occur in Newtonian flow. 
The terminal regime at stronger flow-strength (large $U^{*}$) flows
manifests as a new constant value in Cs-$U^{*}$ curve. 
The Cs-$U^{*}$ curve shown in Figure~\ref{fig:cs_to_ustar}(b)
looks qualitatively similar to the steady state viscosity 
curve of the Moore model flow in Figure~\ref{fig:vis_ss_of_model}. 
However, their mechanisms are different. 
In the case of the steady viscosity,
the thixotropic model fluid structure can be completely broken at strong shear flow, 
because there is no structure-convection effect
under homogenous flow condition as $\nabla \lambda = 0$.
Therefore, the normalized steady state viscosity 
converges to the value of $\xi=0.00990$.
Yet, the terminal value of Cs is not equal to $\xi$. 
Rather, it converges to 0.15 that is almost 15 times larger than 
the pre-expected $\mathrm{Cs}_{\mathrm{min}}=\xi$. 
This is because the second plateau in Cs-$U^{*}$ curve 
is a consequence of the compensation
of structure breakdown by structure convection.

\section{Effect of other dimensionless parameters}
\label{sec:other_effect}

In the previous section, 
we focused on how $\lambda(\bx)$ changes 
depending on $U^{*}$ with regards to equilibrium of three competing factors. 
In this procedure, we employed the drag coefficient Cs
for a quantitative discussion on thixotropy in our model problem. 
In this section, we use the same framework 
to study the effect of other dimensionless numbers.

\subsection{Destruction parameter}
\label{sub_sec:destruction}

\begin{figure}
\begin{center}
\begin{tikzpicture}
\draw (0, 0) node[inner sep=0]{
\includegraphics[width=1.0\textwidth,trim=50 322 20 300 mm, clip=true]
{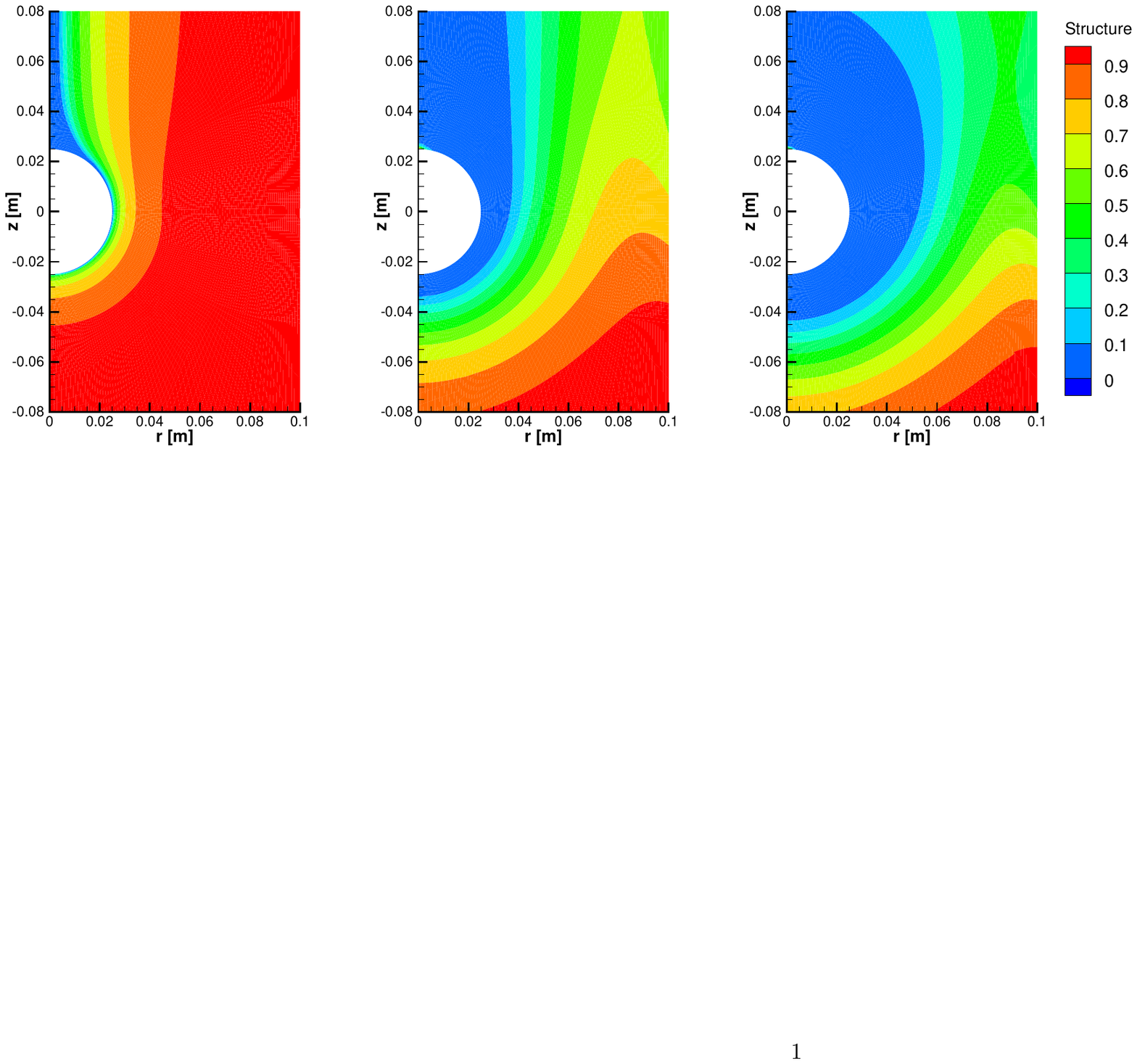}}; 
\draw (-5.65, 3.1) node {(\textbf{a}) $k_d=0.1$};  
\draw (-0.5, 3.1) node {(\textbf{b}) $k_d=1.0$};  
\draw (4.5, 3.1) node {(\textbf{c}) $k_d=8.0$};  
\end{tikzpicture}
\end{center}
\caption{The shape of $\lambda(\bx)$-solution with respect to $k_d$
at $U^{*}=50$. 
$k_d$ determines the equilibrium between the shear-induced breakdown 
and the structure convection for large $U^{*}$ values.
As $k_d$ increases, the fully broken structure (blue region)
around the sphere expands to far field.}
\label{fig:lambda_for_kd}
\end{figure}

\begin{figure}
\begin{center}
\begin{tikzpicture}
\begin{loglogaxis}
[ 
ylabel={Cs \small $\left(=\frac{\cD}{K6\pi (\eta_\infty+\eta_{str}) a U} \right)$},
xlabel={$U^{*}$ \small $\left(=\frac{k_d}{k_a a}U \right)$},
xmin=10^(-3),
xmax=0.5*10^6,
ymin=5*10^(-3), 
ymax=1.5,
legend pos=outer north east,
width=0.6\textwidth,
]
\addplot[mark=*, mark options ={fill=white}, ultra thick, color=blue, mark size =2.0]
table[x expr=\thisrowno{0},y expr=\thisrowno{1},y error expr=\thisrowno{2}] 
{Figures/Cs_Wi_a025_kdpoint1.txt};
\addlegendentry[text width=7.0em]{$k_d=0.1$}
\addplot[mark=*, mark options ={fill=white}, ultra thick, color=red, mark size =2.0]
table[x expr=\thisrowno{0},y expr=\thisrowno{1},y error expr=\thisrowno{2}] 
{Figures/Cs_Wi_a025_revised.txt};
\addlegendentry[text width=7.0em]{$k_d=1$}
\addplot[mark=*, mark options ={fill=white},ultra thick, color=green, mark size =2.0]
table[x expr=\thisrowno{0},y expr=\thisrowno{1},y error expr=\thisrowno{2}] 
{Figures/Cs_Wi_a025_kd8.txt};
\addlegendentry[text width=7.0em]{$k_d=8$}
\addplot[mark=square, mark options ={fill=white}, thick, color=brown, mark size =2.0]
table[x expr=\thisrowno{0},y expr=\thisrowno{1},y error expr=\thisrowno{2}] 
{Figures/Cs_Wi_Cross_revised.txt};
\addlegendentry[text width=7.0em]{Cross Model}
\addplot[dashed, thick, color=black]
table[x expr=\thisrowno{0},y expr=\thisrowno{1},y error expr=\thisrowno{2}] 
{Figures/Cs_Wi_a025_Newton3.txt};
\node at (axis cs:1, 0.013)
{$\mathrm{Cs}_{\mathrm{min}}=\xi=0.00990$};
\end{loglogaxis}
\end{tikzpicture}
\end{center}
\caption{The effect of the destruction parameter $k_d$ on the resistance. 
As $k_d$ increases, fully broken region expands around the
sphere and hence $\eta_e$ decreases. 
A finite value of $k_d$ distinguishes Moore thixotropy
model from Generalized Newtonian model for $U^{*}\gg1$.
}
\label{fig:Cs_ustar_kds}
\end{figure}
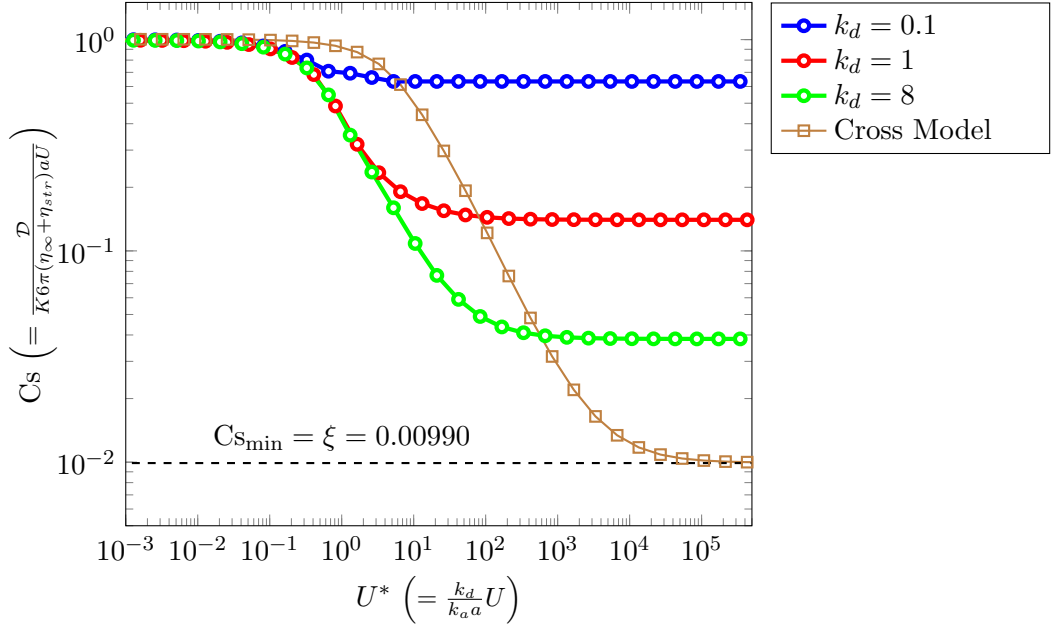

Here, we examine the effect of the destruction parameter $k_d$ 
on microstructure profile $\lambda(\bx)$
and Cs of the sphere. 
Shown in Figure~\ref{fig:lambda_for_kd} is
$\lambda(\bx)$ from additional simulations that are conducted with $k_d=0.1$ and $8$ 
at $U^{*}=50$.
The confinement ratio $a/R$ remained the same as before.
Herein, the boundary condition $U$ (falling velocity) 
is modified accordingly to obtain the same $U^{*}$ condition 
for thixotropic fluids with different $k_d$ values.
As $k_d$ increases, fully broken structure 
(blue region) expands around the sphere. 
The increase in $k_d$ shifts the balance,
which determines the qualitative shape of $\lambda(\bx)$,
toward more broken structure 
in the flow domain by intensifying the effect of shear-induced breakdown.
Cs-$U^{*}$ curves for 3 different $k_d$ values are shown 
in Figure~\ref{fig:Cs_ustar_kds}.
The variations in the shape of $\lambda(\bx)$ 
is well-reflected in the Cs-$U^{*}$ curve, 
the analysis of which enables more quantitative explanation of
the balance shift induced from the increase in $k_d$.
At small $U^{*}$, there exists no remarkable difference in Cs,
which is attributed to the fact that thixotropic behavior in this regime
is dominated by Brownian structure recovery. 
As $U^{*}$ increases, the shear-induced breakdown 
and the convection become
the two dominating factors that determine Cs. 
At large $U^{*}$, the larger $k_d$ results the lower Cs 
in the terminal regime,
indicating that the balance between breakdown 
and convection is achieved when the structure is more broken. 

Comparison of thixotropic fluid with Generalized Newtonian Fluid (GNF), 
which takes a form of $\btau=\btau(\bgamma)$, 
provides a useful insight on the role of destruction parameter $k_d$. 
The brown curve with square symbol in Figure~\ref{fig:Cs_ustar_kds} 
represents the Cs-$U^{*}$ curve of Cross model \cite{Cross:1965}
\begin{equation}
\btau_{\mathrm{cross}}(\dot{\gamma}_s)=
\eta_{\mathrm{cross}}(\dot{\gamma}_s) \bgamma=
\left[\eta_\infty+\frac{\eta_{str}}{1+(k_c\dot{\gamma})^{n}} \right]
\bgamma.
\end{equation}
Cross fluid has been often compared to
the Moore thixotropy model \cite{Barnes:1997,Barnes:1999},
since it has the same steady-state viscosity $\eta_{\mathrm{ss}}$ 
at simple shear flow if $n = 1$, and $k_c=t_c(= k_d/k_a)$. 
For a Cross fluid, we characterize the flow strength $U^{*}$
\begin{equation}
U^{*}_{\mathrm{cross}}= \frac{k_cU}{a}.
\end{equation}
The numerical scheme used for Cross fluid simulation is similar to 
the current numerical scheme. 
In this case, 
we do not need to consider the structure-kinetics equation,
and Picard iterations proceed as follow,
\begin{align}
-\nabla p^{k+1} + \nabla\cdot\left[ \eta_{\mathrm{cross}}(\dgamma^k)\, \bu^{k+1}\right] &= 0,\\
\nabla \cdot\bu^{k+1} &=0.
\end{align}
More details on numerical procedures for GNF fluid are
available elsewhere \cite{JK:2019}. 
The Cs of Cross fluid at the terminal regime $U^{*}\gg 1$ 
is found as $\mathrm{Cs}_{\mathrm{min}}=0.00990$. 
This can be accounted for the absence of 
convection effect in Cross fluid;
GNF model assumes that the fluid rheology
immediately responses to applied shear field 
and thus the convection is ignored. 
Consequently, fluid elements at large $U^{*}$ exist 
with fully broken state $\lambda \approx 0$ 
and viscosity $\eta(\bx)$ over all domain can converge to $\eta_\infty$. 

It is worth noting that the Cs of thixotropic fluid at large $U^{*}$ converges 
to that of Cross fluid as $k_d$ increases. 
This is because the increase in $k_d$ induces fluid elements 
to respond more instantly to the applied shear-field. 
Therefore, when $k_d \to \infty$,
the convection effect becomes insignificant in thixotropic fluid
as in Cross fluid, which is intrinsically lacking the convection.
This causes Cross fluid-like behavior of thixotropic fluid for large $k_d$.

\subsection{The viscosity ratio}
\label{sub_sec:xi}

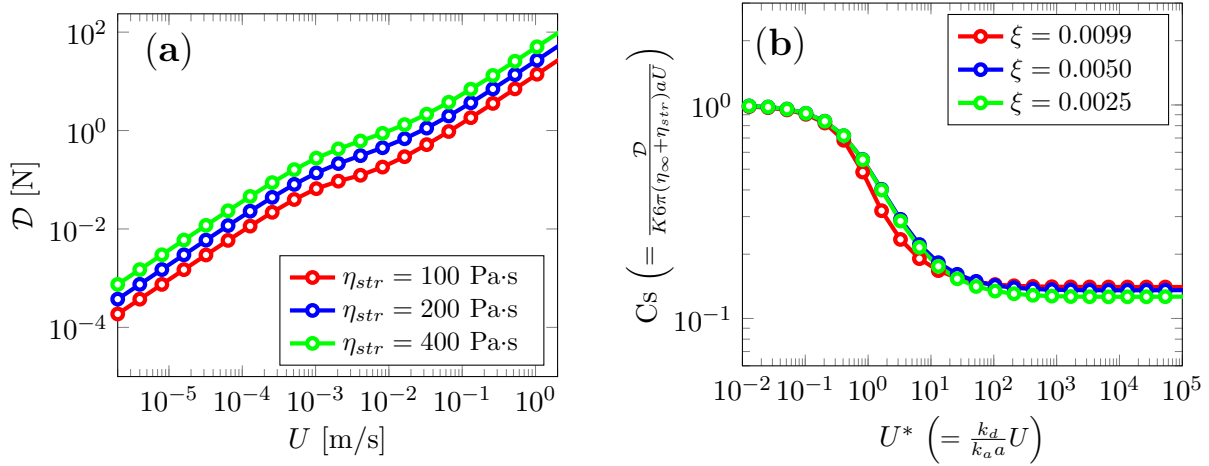
\begin{figure}
\begin{subfigure}{0.45\textwidth} 
\begin{center}
\begin{tikzpicture}
\begin{loglogaxis}
[ 
ylabel={$\cD$ [N]},
xlabel={$U$ [m/s]},
xmin=2*10^(-6),
xmax=2,
ymin=10^(-5), 
width=1.0\textwidth,
legend pos=south east,
]
\node at (axis cs:0.00001, 40) {\Large (\textbf{a})};
\addplot[mark=*,mark options ={fill=white}, ultra thick, color=red, mark size =2.0]
table[x expr=\thisrowno{0},y expr=\thisrowno{1},y error expr=\thisrowno{2}] 
{Figures/D_U_a025.txt};
\addlegendentry[text width=6.5em]{\small $\eta_{str}=100$ Pa$\cdot$s}
\addplot[mark=*,mark options ={fill=white}, ultra thick, color=blue, mark size =2.0]
table[x expr=\thisrowno{0},y expr=\thisrowno{1},y error expr=\thisrowno{2}] 
{Figures/D_U_a025_etaSTR_200.txt};
\addlegendentry[text width=6.5em]{\small $\eta_{str}=200$ Pa$\cdot$s}
\addplot[mark=*,mark options ={fill=white}, ultra thick, color=green, mark size =2.0]
table[x expr=\thisrowno{0},y expr=\thisrowno{1},y error expr=\thisrowno{2}] 
{Figures/D_U_a025_etaSTR_400.txt};
\addlegendentry[text width=6.5em]{\small $\eta_{str}=400$ Pa$\cdot$s}
\end{loglogaxis}
\end{tikzpicture}
\end{center}
\end{subfigure} \hspace{0.03\textwidth}
\begin{subfigure}{0.45\textwidth} 
\begin{center}
\begin{tikzpicture}
\begin{loglogaxis}
[ 
ylabel={Cs \small $\left(=\frac{\cD}{K6\pi (\eta_\infty+\eta_{str}) a U} \right)$},
xlabel={$U^{*}$ \small $\left(=\frac{k_d }{k_a a} U \right)$},
xmin=10^(-2),
xmax=10^(5),
ymin=0.6*10^(-1), 
ymax=3,
legend pos=north east,
width=1.0\textwidth,
]
\node at (axis cs:0.04, 1.9) {\Large (\textbf{b})};
\addplot[mark=*,mark options ={fill=white}, ultra thick, color=red, mark size =2.0]
table[x expr=\thisrowno{0},y expr=\thisrowno{1},y error expr=\thisrowno{2}] 
{Figures/Cs_Wi_a025_revised.txt};
\addlegendentry[text width=5.0em]{\small $\xi=0.0099$}
\addplot[mark=*,mark options ={fill=white}, ultra thick, color=blue, mark size =2.0]
table[x expr=\thisrowno{0},y expr=\thisrowno{1},y error expr=\thisrowno{2}] 
{Figures/Cs_Wi_a025_etastr_200.txt};
\addlegendentry[text width=5.0em]{\small $\xi=0.0050$}
\addplot[mark=*,mark options ={fill=white}, ultra thick, color=green, mark size =2.0]
table[x expr=\thisrowno{0},y expr=\thisrowno{1},y error expr=\thisrowno{2}] 
{Figures/Cs_Wi_a025_etastr_400.txt};
\addlegendentry[text width=5.0em]{\small $\xi=0.0025$}
\end{loglogaxis}
\end{tikzpicture}
\end{center}
\end{subfigure}
\caption{The effect of the structural viscosity $\eta_{str}$ on (\textbf{a}) $\cD$-$U$ curve 
and corresponding (\textbf{b}) Cs-$U^{*}$. 
The increase in $\eta_{str}$ shifts the $\cD$-$U$ curve upward, 
whereas $\cD$-$U$ remains relatively same because of 
the normalization factor $(\eta_\infty+\eta_{str})$ in the definition of Cs.
In (\textbf{a}), it is shown that even at lower $\xi$ values
the transition from full structure viscosity to lower viscosity after a critical shear-rate
still remains gradual in the current linear thixotropy model. 
}
\label{fig:xi_effect}
\end{figure}

In this section, we focus on the effect of the viscosity ratio $\xi$. 
The analogy between the purely-viscous Moore thixotropy model and 
Generalized Newtonian model naturally extends to a discussion 
on the capability of the Moore thixotropy model 
for a thixo-viscoplastic materials. 
Ideally, the effective viscosity of an viscoplastic fluid
is an infinite value. For example, the viscosity $\eta_{B}$ of 
Bingham plastic fluid with two model parameter $\tau_y$ and $\eta_\infty$,
\begin{equation}
\eta_B = \frac{\tau_y}{\dgamma} + \eta_\infty.
\end{equation}
diverges to infinity in the limit of $\dgamma \to 0$.
In numerical investigations, however, the non-elastic 3D versions 
of simple yield stress models are often employed 
with a regularized finite-value of viscosity \cite{Mitsoulis:2017,Balamforth:2014}. 
In case of Moore thixotropy, this corresponds to a case where
$\eta_{str}$ is far larger than $\eta_{\infty}$ (thus, $\xi \to 0$).
Given the similarity, for low values of $\xi$, it can be conjectured that 
the the current pure-viscous thixotropic model can achieve 
what simple viscoplastic models predict.\\
\indent Additional simulations are conducted with different ($\eta_{str}$=200, 400 Pa$\cdot$s) values,
while other model parameters are fixed as given in Table~\ref{tab:model_parameter}.
These cases correspond to $\xi$=0.0050 and 0.0025. 
Figure~\ref{fig:xi_effect} summarizes the results both in $\cD$-$U$ and Cs-$U^{*}$ forms. 
As seen in Figure~\ref{fig:xi_effect}, the Cs-$U^{*}$ curves remain 
relatively same due to the normalization factor $(\eta_\infty+\eta_{str})$ 
in the definition of Cs. 
When it comes to the $\cD$-$U$ curve, 
the increase in $\eta_{str}$ shifts the curve upward
due to the effect of increased viscosity around the sphere. 
Yet, even at lower $\xi$ values, 
the transition from fully structured viscosity to lower viscosity after a critical shear-rate
still remains gradual. 
This is contrast to what typical regularized viscoplastic models predict; 
in these models, a dramatic viscosity drop after a critical shear-rate
manifests as a plateau in $\cD$-$U$ curve \cite{JK:2019}. 
This is owing to the viscosity regularization terms,
which are intentionally designed to mimic yielding processes. 
It is conjectured that 
the current linear model form, 
either structure-kinetics (\ref{e:Toorman_kinetic})
or viscosity (\ref{e:Moore_viscosity}),  
should be modified to have exponential 
dependence on $\lambda$ \cite{Ardakani:2011,Alexandrou:2009}
in order to achieve what simple viscoplastic model predicts.
Yet, this extension shares the same underlying assumption
that the materials exhibits no elasticity. 
In addition, this naive approach is fundamentally limited 
when it comes to predicting the important issue of 
the presence of anisotropic yield stress \cite{Thompson:2018}.
The interested reader on this issues
is referred to additional reference 
where rheological models consider elastic and thixotropic response
to a combination of shear and extensional deformations \cite{Tsamopoulos:2016,Dinkgreve2017}
and computational implementation of these models in real flow scenarios \cite{new_cite4,new_cite5}.

\subsection{Confinement effect}
\label{sec_confinement}

\begin{figure}
\begin{center}
\begin{tikzpicture}
\draw (0, 0) node[inner sep=0]{
\includegraphics[width=1.0\textwidth,trim=50 322 20 300 mm, clip=true]
{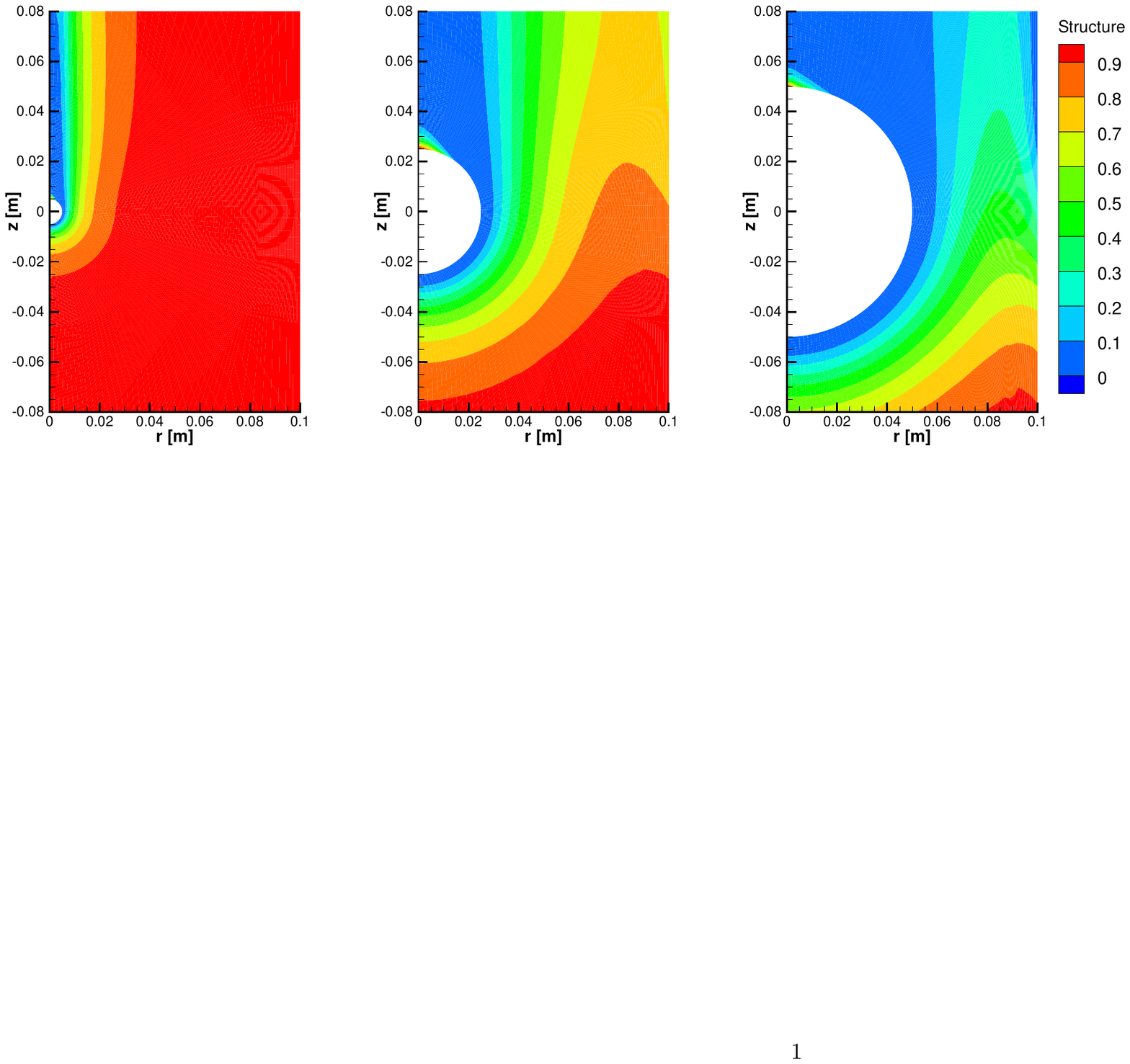}}; 
\draw (-5.65, 3.1) node {(\textbf{a}) $a/R=0.05$};  
\draw (-0.5, 3.1) node {(\textbf{b}) $a/R=0.25$};  
\draw (4.7, 3.1) node {(\textbf{c}) $a/R=0.5$};  
\end{tikzpicture}
\end{center}
\caption{Solutions of structure $\lambda$ 
for different confinement
ratio $a/R$ at fixed $U^{*}=50$. 
As $a/R$ increases, the balance between shear-break 
and the structure convection is achieved 
with more broken structures in the vicinity of sphere. 
When $a/R=0.5$, the wall interaction starts to appear 
with the broken structures near the side wall.}
\label{fig:lambda_for_atoR}
\end{figure}

Geometry is another important factor that significantly contributes to 
a non-homogeneous flow of thixotropic fluid. 
In this section, we study how a geometrical factor 
modifies the competition of the three factors. 
Provided that the length of confining cylinder $L$ is sufficiently large enough, 
the geometrical condition of the our model problem is uniquely characterized 
by the confinement ratio $a/R$.
Accordingly, additional simulations are conducted with $a/R = 0.05, 0.25$ and $0.5$,
while other model parameters are fixed as given in Table~\ref{tab:model_parameter}.
Again, the falling velocity $U$ is modified to obtain the same $U^{*}$ condition.  
Shown in Figure~\ref{fig:lambda_for_atoR} is the structure profile $\lambda(\bx)$ 
observed in three different geometry at $U^{*}=50$.
The value is located in the end of the transient regime in Figure~\ref{fig:cs_to_ustar}(b).
It shows significant difference in $\lambda(\bx)$ at different $a/R$ values. 
In the case of the small sphere ($a/R = 0.05$),
the bulk of fluid elements passes through the 
large area between the sphere and wall
and thus the fluid is relatively less sheared 
compared to the large sphere case. 
Thus, shear-induced breakdown becomes insignificant
whereas the convecting fresh full-structured elements 
easily refill the space around the sphere. 
As a result, the thixotropic fluid sustains a structured state 
as shown in Figure~\ref{fig:lambda_for_atoR}(a). 
On the contrary, if $a/R$ increases, 
the thixotropic fluids elements become highly sheared,
passing through a narrower area between sphere and wall.
This makes the shear-induced breakdown predominant over refilling of fresh element. 
Consequently, the balance between shear-induced breakdown and 
the convection is achieved at more broken structure as shown in 
Figure~\ref{fig:lambda_for_atoR}(c).

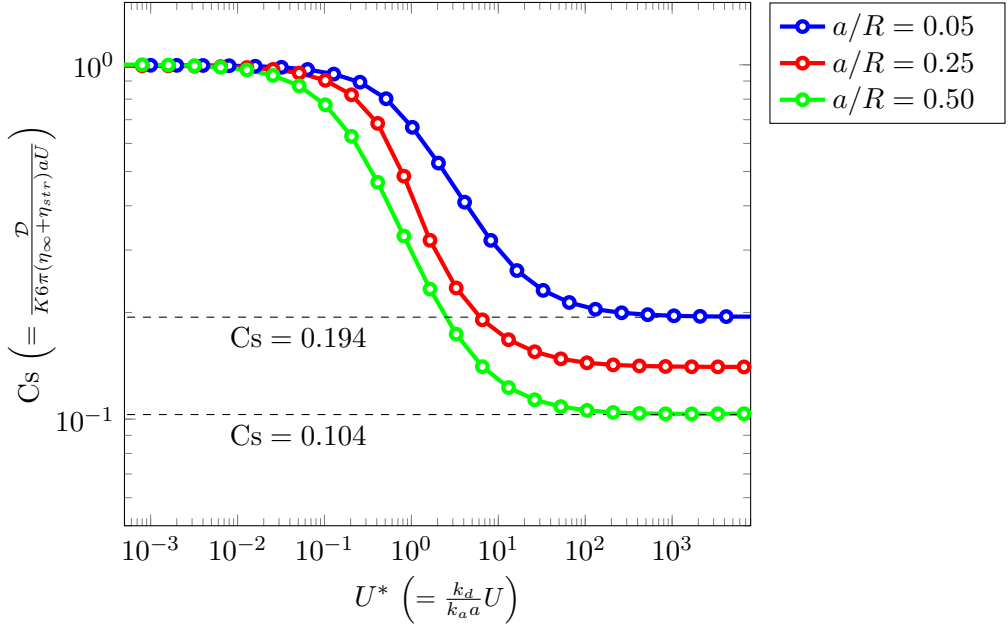
\begin{figure}
\begin{center}
\begin{tikzpicture}
\begin{loglogaxis}
[ 
ylabel={Cs \small $\left(=\frac{\cD}{K6\pi (\eta_\infty+\eta_{str}) a U} \right)$},
xlabel={$U^{*}$ \small $\left(=\frac{k_d}{k_a a}U  \right)$},
xmin=0.5*10^(-3),
xmax=8 * 10^3,
ymin=0.05, 
ymax=1.5,
width=0.6\textwidth,
legend pos=outer north east,
]
\addplot[mark=*, mark options ={fill=white},ultra thick, color=blue, mark size =2.0]
table[x expr=\thisrowno{0},y expr=\thisrowno{1},y error expr=\thisrowno{2}] 
{Figures/Cs_Wi_a005_revised.txt};
\addlegendentry[text width=5.5em]{$a/R=0.05$}
\addplot[mark=*, mark options ={fill=white},ultra thick, color=red, mark size =2.0]
table[x expr=\thisrowno{0},y expr=\thisrowno{1},y error expr=\thisrowno{2}] 
{Figures/Cs_Wi_a025_revised.txt};
\addlegendentry[text width=5.5em]{$a/R=0.25$}
\addplot[mark=*, mark options ={fill=white},ultra thick, color=green, mark size =2.0]
table[x expr=\thisrowno{0},y expr=\thisrowno{1},y error expr=\thisrowno{2}] 
{Figures/Cs_Wi_a05_revised.txt};
\addlegendentry[text width=5.5em]{$a/R=0.50$}
\addplot[mark=none, black, dashed] coordinates {(10^(-4),0.194) (10^4,0.194)};
\node at (axis cs:0.05, 0.17)
{$\mathrm{Cs}=0.194$};
\addplot[mark=none, black, dashed] coordinates {(10^(-4),0.103) (10^4,0.103)};
\node at (axis cs:0.05, 0.09)
{$\mathrm{Cs}=0.104$};
\end{loglogaxis}
\end{tikzpicture}
\end{center}
\caption{The variation of Cs-$U^{*}$ curve at different confinement 
ratio ($a/R$). The effect of confinement is negligible at small $U^{*}$
At larger $a/R$, Cs drops much faster to lower plateau 
as $U^{*}$ increases.
Such effect is not confined to the terminal regime
but greatly changes the shape of the transient regime,
contrast to the effect of the destruction parameter $k_d$. 
}
\label{fig:cs_to_atoR}
\end{figure}

The aspect of $\lambda(\bx)$ transition with respect to $a/R$ variation
is reflected in Cs-$U^{*}$ curve in Figure~\ref{fig:cs_to_atoR} as well.
Here, we remind that our definition of Cs~(\ref{e:Cs}) 
takes into account the wall interaction factor,
extracting the effect of  thixotropy effect on resistance. 
The overall trend of three curves in Figure~\ref{fig:cs_to_atoR}
is similar; they are all characterized by a sigmoid shape with two horizontal asymptotes. 
At large $a/R$, however, Cs drops much faster toward a lower plateau. 
The bigger and faster drop of Cs with respect to the increase in $U^{*}$
is correlated to the balance achieved at more broken structure 
due to predominant shear-induced breakdown. 

Considering how the variation of $a/R$ affects 
both the transient regime of Cs-$U^{*}$ curve and the terminal regime, 
the geometrical factor distinguishes itself 
from the destruction parameter $k_d$, 
the effect of which is mostly limited to the terminal regime. 
This suggests that a behavior of thixotropic fluid 
is not a sole result of the intrinsic material property, 
but that of involved interplay between 
intrinsic property and 
extrinsic geometrical factors of a flow.   
The implication is that one can accomplish 
a desired flow condition by the tuning flow geometry, 
which can be often easier than modifying
the intrinsic property of a material.

\section{Nonlinear thixotropic models}
\label{sec:nonlinear}

In this section, we examine how an addition 
of nonlinearity to the Moore thixotropy affects 
the non-homogenous flow.  
Although the Moore thixotropy model has been 
considered as a simple but effective model for various kinds
of thixotropic fluids, it is sometimes inadequate
to correctly describe a more complicated 
thixotropic behavior. 
Recent thixotropic modeling works aim to incorporate them by replacing 
explicit model-form with an implicit form \cite{Renardy:2016,Stephanou:2018}.

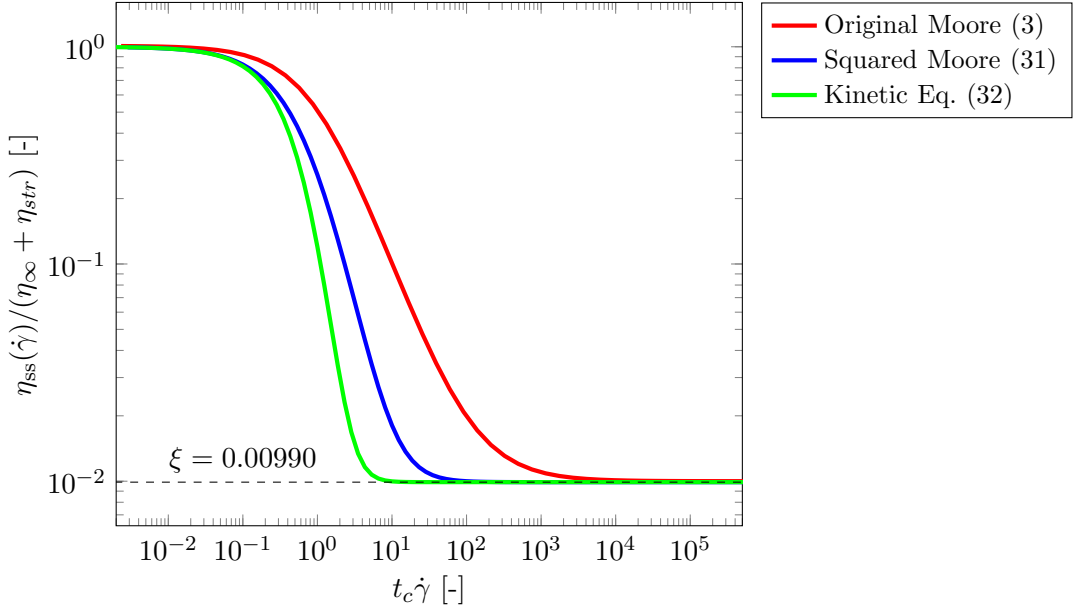
\begin{figure}
\begin{center}
\begin{tikzpicture}
\begin{loglogaxis}
[ 
ylabel={$\eta_{\mathrm{ss}} (\dot\gamma) /(\eta_\infty+\eta_{str})$ [-]},
xlabel={$t_c \dot{\gamma}$ [-]},
xmin=2*10^(-3),
xmax=5*10^5,
width=0.6\textwidth,
legend pos=outer north east,
]
\addplot[no marks, ultra thick, color=red, mark size =2.0]
table[x expr=\thisrowno{0},y expr=\thisrowno{1},y error expr=\thisrowno{2}] 
{Figures/vis_ss_of_model.txt};
\addlegendentry[text width=8.1em]{
\small Original Moore~(\ref{e:Moore_viscosity})}
\addplot[no marks, ultra thick, color=blue, mark size =2.0]
table[x expr=\thisrowno{0},y expr=\thisrowno{1},y error expr=\thisrowno{2}] 
{Figures/vis_ss_of_viscosity_square.txt};
\addlegendentry[text width=8.1em]{
\small Squared Moore~(\ref{e:sq_Moore})}
\addplot[no marks, ultra thick, color=green, mark size =2.0]
table[x expr=\thisrowno{0},y expr=\thisrowno{1},y error expr=\thisrowno{2}] 
{Figures/vis_ss_of_rate_square.txt};
\addlegendentry[text width=8.1em]{
\small Kinetic Eq.~(\ref{e:newToorman})}
\addplot[mark=none, black, dashed] coordinates {(10^(-4),0.00990) (10^6,0.00990)};
\node at (axis cs:0.1, 0.0125)
{$\xi=0.00990$};
\end{loglogaxis}
\end{tikzpicture}
\end{center}
\caption{Steady shear viscosity $\eta_{\mathrm{ss}}$ of nonlinear thixotropic
model considered in Section~\ref{sec:nonlinear}.
The $k_{d2}$ in structure-kinetics Eq.~(\ref{e:newToorman}) is set as 1.0 [s], 
while the remaining model parameters are as in Table~\ref{tab:model_parameter}. 
The structural sensitivity to the applied shear-rate (\ref{e:newToorman}) 
is more significant than the sensitivity of the viscosity to microstructural change (\ref{e:sq_Moore}). 
}
\label{fig:vis_ss_of_second_order_model}
\end{figure}

Within the purely-viscous thixotropic model context, 
a variety of modifications of Moore thixotropic
model have also been employed for a more accurate 
description of sophisticated thixotropic behavior~\cite{Mewis:2011,Mewis:2009}. 
The general modification is to include nonlinearity by considering high-order
polynomial terms to either the constitutive equation
\begin{equation}
\btau(\lambda) = \eta_\infty \bgamma + \left( \eta_{str,1} \lambda 
+\eta_{str,2} \lambda^2 + \hdots \right) \bgamma,
\end{equation} 
or the structure kinetic equation,
 \begin{equation}
\frac{d\lambda}{dt}=\left(k_{a,1} [1-\lambda] + k_{a,2} [1-\lambda]^{2} + \hdots \right)
- \left(k_{d1} \dgamma \lambda  + k_{d2} \dgamma^2  \lambda^{2}+ \hdots \right).
\end{equation} 
Because the additional high-order terms provide more
degree of freedom to fit steady-shear and transient shear data simultaneously, 
these forms have been adopted for modeling 
slurry fuels \cite{Lin:1985,Pinder:1964} and 
semisolid metal suspensions \cite{Burgos:2001}.
These additional material parameters cannot 
be directly related to the micromechanics  of real material; 
the structure-kinetic method is phenomenological modeling 
in nature \cite{Mewis:2009,Stephanou:2018}. 
In general, the use of extra high order terms should be 
penalized, unless they are legitimately supported from 
existing rheological data \cite{Freund:2015, Jeffreys:1939}. 
Thus, we investigate two simple cases
where nonlinearity is implemented by an addition 
of a second-order term respect to 
$\lambda$ or $\dgamma$. 
It should be noted that these modified model forms are still purely viscous fluid
and only consider shear-dependent rheology.

\begin{figure}
\begin{subfigure}{0.45\textwidth} 
\begin{center}
\begin{tikzpicture}
\begin{loglogaxis}
[ 
ylabel={$\cD$ [N]},
xlabel={$U$ [m/s]},
xmin=2*10^(-6),
xmax=50,
ymin=10^(-4), 
ymax=200,
width=1.0\textwidth,
legend style={legend pos=south east}
]
\node at (axis cs:0.00001, 50) {\Large (\textbf{a})};
\addplot[mark=*, mark options ={fill=white},ultra thick, color=red, mark size =2.0]
table[x expr=\thisrowno{0},y expr=\thisrowno{1},y error expr=\thisrowno{2}] 
{Figures/D_U_a025.txt};
\addlegendentry[text width=8.1em]{
\small Original Moore~(\ref{e:Moore_viscosity})}
\addplot[mark=*, mark options ={fill=white},ultra thick, color=blue, mark size =2.0]
table[x expr=\thisrowno{0},y expr=\thisrowno{1},y error expr=\thisrowno{2}] 
{Figures/D_U_a025_Sq_Moore_revised.txt};
\addlegendentry[text width=8.1em]{
\footnotesize Squared Moore~(\ref{e:sq_Moore})}
\node[rotate=42] at (axis cs:0.01, 3)
{\scriptsize $\cD_N=K 6 \pi  [\eta_\infty + \eta_{str}] a U$};
\addplot[dashed, thick, color=black]
table[x expr=\thisrowno{0},y expr=\thisrowno{1},y error expr=\thisrowno{2}] 
{Figures/Newton_D_U_a025.txt};
\end{loglogaxis}
\end{tikzpicture}
\end{center}
\end{subfigure} \hspace{0.03\textwidth}
\begin{subfigure}{0.45\textwidth} 
\begin{center}
\begin{tikzpicture}
\begin{loglogaxis}
[ 
ylabel={Cs \small $\left(=\frac{\cD}{K6\pi (\eta_\infty+\eta_{str}) a U} \right)$},
xlabel={$U^{*}$ \small $\left(=\frac{k_d}{k_a a} U \right)$},
xmin=0.5*10^(-3),
xmax=10^3,
ymin=0.006, 
ymax=3.5,
width=1.0\textwidth,
]
\addplot[mark=*, mark options ={fill=white},ultra thick, color=red, mark size =2.0]
table[x expr=\thisrowno{0},y expr=\thisrowno{1},y error expr=\thisrowno{2}] 
{Figures/Cs_Wi_a025_revised.txt};
\addplot[mark=*, mark options ={fill=white},ultra thick, color=blue, mark size =2.0]
table[x expr=\thisrowno{0},y expr=\thisrowno{1},y error expr=\thisrowno{2}] 
{Figures/Cs_Wi_Sq_Moore_revised.txt};
\addplot[dashed, thick, color=black]
table[x expr=\thisrowno{0},y expr=\thisrowno{1},y error expr=\thisrowno{2}] 
{Figures/Cs_Wi_a025_Newton3.txt};
\addplot[mark=none, black, dashed] coordinates {(10^(-4),0.094) (10^3,0.094)};
\node at (axis cs:1, 0.075)
{$\mathrm{Cs}=0.094$};
\node at (axis cs:1, 0.014)
{$\mathrm{Cs}_\mathrm{min}=\xi$};
\node at (axis cs:0.0019, 2.0) {\Large (\textbf{b})};
\end{loglogaxis}
\end{tikzpicture}
\end{center}
\end{subfigure}
\caption{
The effect of the modified viscosity function (\ref{e:sq_Moore})
on the (\textbf{a}) $\cD$-$U$ curve and (\textbf{b}) Cs-$U^{*}$ curve  
is plotted. 
The nonlinearity implemented in the viscosity 
function~(\ref{e:sq_Moore}) is still governed by the balance between 
structure breakdown and convection at large $U^{*}$.
}
\label{fig:Cs_Moore_sq}
\end{figure}
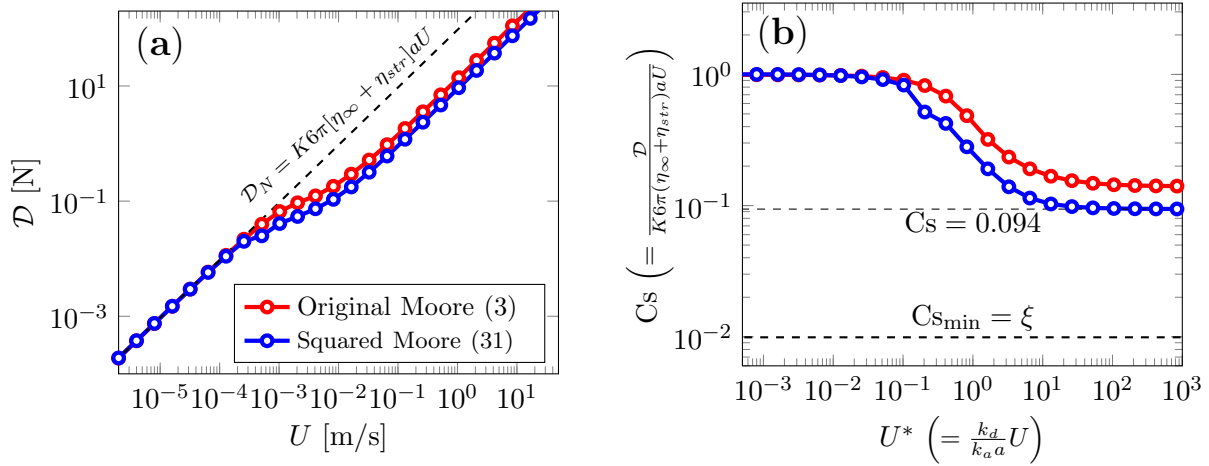

\begin{figure}
\begin{center}
\begin{tikzpicture}
\begin{loglogaxis}
[ 
ylabel={$\cD$ [N]},
xlabel={$U$ [m/s]},
xmin=10^(-5),
xmax=10^3,
ymin=10^(-4), 
ymax=0.5*10^(4),
width=0.6\textwidth,
legend style={legend pos=south east}
]
\addplot[mark=*, thick, mark options ={fill=white}, ultra thick,color=red, mark size =2.0]
table[x expr=\thisrowno{0},y expr=\thisrowno{1},y error expr=\thisrowno{1.5}] 
{Figures/D_U_a025.txt};
\addlegendentry[text width=8.1em]{
\small Original Moore~(\ref{e:Moore_viscosity})}
\addplot[mark=*, thick, mark options ={fill=white}, ultra thick,color=green, mark size =2.0]
table[x expr=\thisrowno{0},y expr=\thisrowno{1},y error expr=\thisrowno{1.5}] 
{Figures/D_U_a025_ksqaure_revised.txt};
\addlegendentry[text width=8.1em]{
\small Kinetics Eq.~(\ref{e:newToorman})}
\addplot[dashed, thick, color=black]
table[x expr=\thisrowno{0},y expr=\thisrowno{1},y error expr=\thisrowno{1.5}] 
{Figures/Newton_D_U_a025.txt};
\node[rotate=42] at (axis cs:0.01, 4)
{$\cD_N=K 6 \pi  (\eta_\infty + \eta_{str})a U$};
\addplot[dashed, thick, color=black]
table[x expr=\thisrowno{0},y expr=\thisrowno{1},y error expr=\thisrowno{1.5}] 
{Figures/Newton_D_U_kdsquare.txt};
\node[rotate=42] at (axis cs:1, 0.3)
{$K 6 \pi \eta_\infty a U$};
\end{loglogaxis}
\end{tikzpicture}
\end{center}
\caption{The resistance prediction 
for the nonlinear structure-kinetics 
equation (\ref{e:newToorman}). 
When the structure is more sensitive 
to applied shear rate than a linear scale, 
shear-induced breakdown effect dominates 
the Brownian recovery and structure convection
as $U^{*}$ increases.
Therefore, $\cD$ converges to Newtonian resistance 
with viscosity $\eta_{\infty}$. }
\label{fig:sqthixo}
\end{figure}
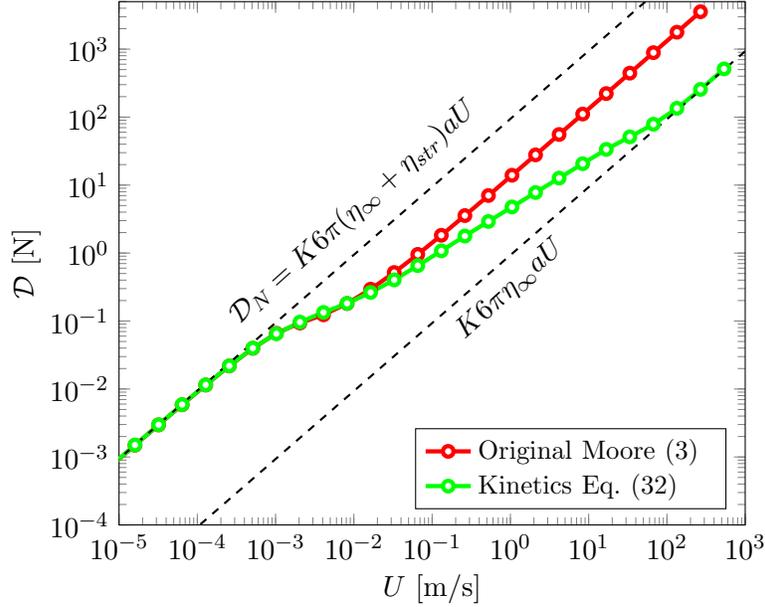

Firstly, we consider a nonlinearity in viscosity function
\begin{equation}
\eta(\lambda) = \eta_\infty + \eta_{str}\lambda^2.
\label{e:sq_Moore}
\end{equation}
instead of Eq.~(\ref{e:Moore_viscosity}).
The other governing equations, including 
the structure-kinetics equation, remain the same.
The steady state shear viscosity $\eta_{ss}$ of
the viscosity model Eq.~(\ref{e:sq_Moore}) is shown with 
the blue curve in Figure~\ref{fig:vis_ss_of_second_order_model}.
It indicates that the viscosity of a material with extra higher order terms
is more sensitive to 
microstructural change than the original is.
In Figure~\ref{fig:Cs_Moore_sq}(a), 
the resistance $\cD$ of a settling sphere in this model fluid 
is compared with that of the Moore viscosity.
Here, the material parameters 
$\{\eta_\infty,\eta_{str},k_d,k_a \}$ of both models are
the same as those summarized in Table~\ref{tab:model_parameter}.
Based on this result, Cs-$U^{*}$ curves of the original Moore model 
and the nonlinear viscosity function (square term) 
are compared in Figure~\ref{fig:Cs_Moore_sq}(b). 
As $U^{*}$ increases, the Cs value 
of the new viscosity model shows a faster 
and larger drop because of the higher viscosity-sensitivity 
to structural breakdown (or decrease in $\lambda$).
Yet, the terminal Cs value (0.095) results in 
the squared model~(\ref{e:sq_Moore})
is still nearly 10 times larger than 
$\mathrm{Cs}_{\mathrm{min}}$.
This is because the results in the nonlinear viscous model
are still governed by the balance between the structural 
breakdown and the structure convection.

Next, we focus on another modification of the Moore model 
where nonlinearity is introduced by the addition of 
a square term to the structure-kinetics equation. 
In this case, the kinetics equation is 
given as 
\begin{equation}
\frac{d\lambda}{dt}=\frac{\partial \lambda}{\partial t} + (\bu \cdot \nabla \lambda) 
=-k_{d1} \dgamma \lambda - k_{d2} \dgamma^2 \lambda+ 
k_a (1-\lambda),
\label{e:newToorman}
\end{equation}
instead of Eq.~(\ref{e:Toorman_kinetic}), 
while the constitutive equation is 
same as the original Eq.~(\ref{e:Moore_viscosity}).
In Figure~\ref{fig:vis_ss_of_second_order_model},
the steady state shear viscosity $\eta_{ss}$ of
the structure-kinetics model~(\ref{e:Toorman_kinetic})
is shown with the green curve in Figure~\ref{fig:vis_ss_of_second_order_model}.
In this case, the homogenous structure solution in 
a simple shear flow changes to 
$\lambda_{ss}=1/(t_c\dot\gamma+(k_{d2}/k_a)\dot{\gamma}^2+1)$
the structural change is more sensitive to
the applied shear-rate. 
As shown in Figure~\ref{fig:vis_ss_of_second_order_model},
The structural sensitivity to the applied shear-rate is more significant 
than the sensitivity of viscosity to microstructural change. 
In Figure~\ref{fig:sqthixo}, 
the resistance $\cD$ calculated from the original Moore model and
the model with nonlinear structural kinetic equation~(\ref{e:newToorman})
are compared. 
In both cases, material parameters 
$\{\eta_\infty,\eta_{str},k_{d1}=k_d,k_a \}$
are given to match values in
Table~\ref{tab:model_parameter}
and the new parameter $k_{d2}$ in this nonlinear model is set as 1.0 [s]. 
Here, we report the simulation result with units (i.e., $\cD$-$U$ curve),
because the strength of shear-induced breakdown 
from the kinetic equation~(\ref{e:newToorman}) is much stronger 
than that of our previous definition of $U^{*}$~(\ref{e:dimensionless_U}); 
while breakdown effect is mostly governed by 
the second-order term $k_{d2}\dgamma^2$ when $U^{*} >1$, 
we cannot compare these effect in the same scale 
because of the absence of the term in the original Moore model. 
As expected from steady shear viscosity (Figure~\ref{fig:vis_ss_of_second_order_model}),
the Moore model with nonlinear structure-kinetic equation 
shows a faster transition in $\cD$-$U^{*}$ curve 
compared to the previous case with nonlinearity viscosity function.
It also shows a more notable difference 
in that $\cD$ finally converges 
to that of Newtonian fluid with viscosity $\eta_{\infty}$
as $\mathrm{U}\to \infty$. 
This is because the strengthened breakdown effect 
becomes completely dominant over other two factors,
and thus the structure convection cannot compensate 
the breakdown effect as $U^{*} \to \infty$.

In summary, adding nonlinearities in either in
the viscosity function 
and the structure kinetic equation work differently. 
Thus, when nonlinear thixotropic models 
are considered, 
one should carefully distinguish 
whether the constitutive equation or the structural change equation 
is more sensitive upon consideration of 
relevant experimental measurement or designing flow process.

\section{Conclusion}
\label{sec:conclusion}

In this work, a non-homogenous flow of
a Moore thixotropic fluid is explored in terms of an interplay between 
intrinsic thixotropy of materials and geometry of flow. 
As an example, the steady thixotropic flow around 
a settling sphere is analyzed by a numerical simulation.
Combined with a typical Stokes flow solver,
we employed an Discontinous Galerkin approximation for 
advection-type structure-kinetics equation. 
The structure profile $\lambda(\bx)$ in the flow
is quantitatively characterized by 
the resistance coefficient Cs and classified 
into three different regimes according to
the normalized velocity $U^{*}$. 
This transition is discussed with regards to the balance of 
three competing factors: Brownian structure 
recovery, shear-induced structure breakdown, 
and the structure convection. 
At small $U^{*}$, thixotropic effect is negligible and 
Newtonian-like behavior is observed, 
since the overall dynamic is overwhelmed by
the Brownian structure recovery. 
As $U^{*}$ increases, the shear-induced breakdown 
and the convection become significant
and the equilibrium of three factors shifts toward
more structure-broken state. 
At large $U^{*}$, the final balance between 
breakdown and the convection is achieved,
since both effects are proportional to $U^{*}$. 
Based on our findings, the effects of a destruction 
parameter and confinement are discussed. 
The destruction parameter $k_d$ determines the 
the shape of the structure solution around the sphere 
for large enough $U^{*}$. 
It has been also shown that a finite value of $k_d$ 
distinguishes a thixotropic model fluid 
from a Generalized Newtonian model in a non-homogenous flow.
Moreover, the analysis on confinement effect 
emphasized the importance of geometrical 
factors in a material process with thixotropy. 
Finally, we investigated how two different
ways of implementing nonlinearity in Moore model 
lead to different descriptions of thixotropic behavior. 

Most of previous experimental and theoretical studies 
have focused on intrinsic properties of material itself, 
and thus assumed a homogeneous flow to provide 
effective modeling approaches for thixotropic behavior. 
By contrast, the current work distinguishes itself from 
the existing studies as it extends a scope to illustrating 
a complex interplay between the intrinsic thixotropy of materials and flow geometry. 
A near future work may include typical intricate examples:
particles with non-spherical shape (e.g, oblates \cite{ftw_oblate}, 
cylinders \cite{ftw_cylinder}, cubes \cite{ftw_cube}), 
surrounding fluids with unsteady motion \cite{ftw_unsteady1,
ftw_unsteady2}, or motion of bubbles \cite{ftw_bubble}.
Understanding these problems will pave a way to realistic 
engineering of thixotropic fluid in practical application scenarios,
e.g. establishing an effective flow-assurance strategy
to ensure continuous flow of production 
crude oils and other industrial products \cite{assurance1,assurance2}.

We close by recognizing the
unresolved fundamental issues of a non-homogenous flow 
of complex fluids around a solid sphere
\cite{Denn:2009,Denn:2011}.
Most of materials show more complex rheological 
characteristics such as non-trivial 
extensional rheology, yield stress,
and TEVP features \cite{Denn:2011,Mendes:2011}.
Therefore, understanding non-homogenous flow of complex fluid, 
which is conjectured to be a result of an interplay between 
intricate material rheology and flow geometry, 
is required for accurate design and control of practical fluid processes. 
Although a purely viscous thixotropic model is considered 
in this paper, our result is meaningful as an initial step to unravel 
the longstanding question of non-homogenous flows of complex fluids.

\section{Acknowledgements}

This research did not receive any specific grant from funding
agencies in the public, commercial, or non-for-profit sectors.

\appendixtitleon
\appendixtitletocon
\begin{appendices}
\section{Code verification}
\label{a:code_verification}
\setcounter{equation}{0}  
\numberwithin{equation}{section}
\setcounter{figure}{0}  
\renewcommand\thefigure{\thesection.\arabic{figure}}   

To verify the numerical scheme used in this work, 
the order of accuracy and convergence are tested. 
Because the exact solution of
the set of governing equations 
(\ref{e:gv1}) to (\ref{e:gv3}) is unknown,
the method of manufactured solution is 
considered \cite{MMS1,MMS2}.
In this framework, a selected
`manufactured solution'~$(\widetilde{\bu},\widetilde{p},\widetilde{\lambda})$
is forced to be the exact solution 
by modifying the source terms, 
which are the right hand side of 
(\ref{e:gv1}), (\ref{e:gv2}), and (\ref{e:gv3}).
For example,
we selected our manufactured solution as 
the Newtonian analytic solution of
a flow around sphere falling with velocity $U$ at infinite space \cite{Happel:1965},
\begin{equation}
\widetilde{u}_{r_s}= U \cos{\theta} \left( 1+ \frac{1}{2} \frac{a^3}{r_s^3}-\frac{3}{2}\frac{a}{r_s}\right), 
\label{e:newton_ur}
\end{equation}
\begin{equation}
\widetilde{u}_{\theta}= U \cos{\theta} \left( 1- \frac{1}{4} \frac{a^3}{r_s^3}-\frac{3}{4}\frac{a}{r_s}\right),
\label{e:newton_utheta}
\end{equation}
\begin{equation}
\widetilde{p}=\frac{3}{2}aU \left(\frac{\cos{\theta}}{r_s^2} \right), 
\label{e:newton_pressure}
\end{equation}
where $r_s$ and $\theta$ are the radial distance 
and the polar angle of
a spherical coordinate system 
with the origin at the center of the sphere.
For $\lambda(\bx)$,
the manufactured solution arbitrarily constructed by
\begin{equation}
\widetilde{\lambda} = r_s \theta.
\label{e:mms_lambda}
\end{equation}
In some part of the computational domain $\Omega$, 
the constructed form $\widetilde{\lambda}$ (\ref{e:mms_lambda}) 
exceeds $\lambda>1$, which the model-form does not allow.
Yet, there is no requirement 
for physical realism in the
manufactured solution for the current purpose 
\cite{MMS3}.  
The form (\ref{e:mms_lambda}) is suitable to test 
all terms in the set of governing equations
and simple enough to determine the analytic form 
of the corresponding source terms. 

During the code verification stage,
we assumed the simplest parameter values 
$\eta_\infty=\eta_{\mathrm{s}}=1$ Pa$\cdot$s 
of the Moore viscosity function (\ref{e:Moore_viscosity}).
To find the analytic expression for  
the corresponding source terms in 
the momentum conservation $\vec{F}=(F_{r_s},F_{\theta})$,
we first substituted (\ref{e:mms_lambda})
into the constitutive equation (\ref{e:constitutive_eq}) 
to evaluate the material stress $\btau$.
After, resubstituting the resulted analytic form of stress $\btau$ 
into the conservation of momentum (\ref{e:gv1}),
we obtained
\begin{equation}
F_{r_s}(r_s,\theta)=\frac{aU \big[ (3a^2 \theta -6 \theta r_s^2) \cos{\theta} + \frac{3}{2}a^2 \sin{\theta}  \big] } {r_s^4},
\end{equation}
\begin{equation}
F_\theta(r_s,\theta) =\frac{aU \big[ \frac{3}{2}(r_s^2-a^2) \cos{\theta} + \frac{3}{2}\theta (a^2-r_s^2)\sin{\theta}  \big] } {r_s^4}.
\end{equation}
The source term of the mass conservation remains zero 
as our choice of $\widetilde{\bu}$ automatically 
satisfies the incompressible constraints, i.e. $\nabla \cdot \widetilde{\bu}=0$. 
Similarly, substituting (\ref{e:newton_ur}), (\ref{e:newton_utheta}) and (\ref{e:mms_lambda})
into the kinematic equation (\ref{e:gv3}), 
we found the corresponding source term $Q$ 
of structure-kinetics equation as
\begin{equation}
\begin{aligned}
Q =&\frac{3}{2} k_d \theta  r_s a U^2 \sqrt{\frac{3 \cos ^2 \theta \left(r^2-a^2\right)^2+a^4 \sin ^2\phi }{r^8}}\\
&+k_a \theta r_s+U \sin \phi  \left(\frac{a^3 U}{4 r_s^3}+\frac{3 a U}{4r_s}-1\right)
+\theta  U \cos \theta \left(\frac{a^3 U}{2 r_s^3}-\frac{3 R U}{2 r_s}+1\right).
\end{aligned}
\end{equation}

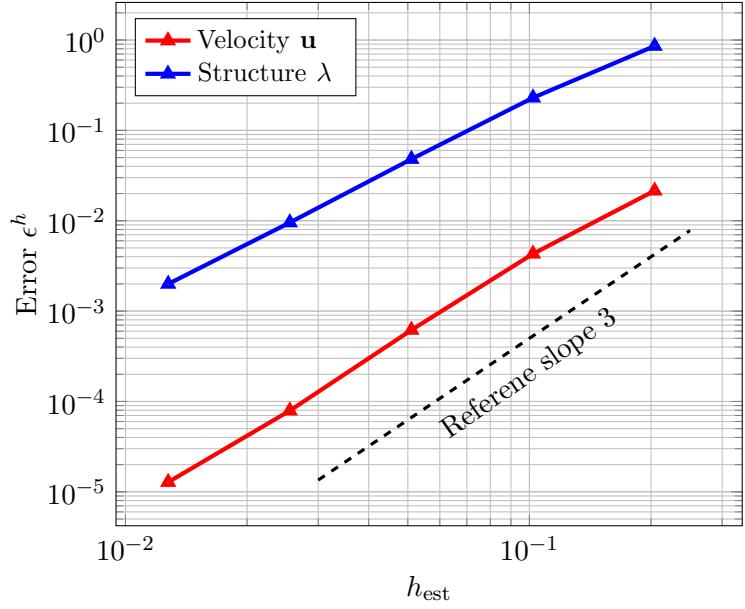
\begin{figure}
\begin{center}
\begin{tikzpicture}
\begin{loglogaxis}
[ 
ylabel={Error $\epsilon^{h}$},
xlabel={$h_{\mathrm{est}}$},
grid=both,
width=0.6\textwidth,
legend style={legend pos=north west}
]
\addplot+[mark=triangle, mark size=2, color=red, ultra thick] 
table[x expr=\thisrowno{0}, y expr=\thisrowno{1}, col sep=space] {Figures/MMF_velocity.txt};
\addlegendentry[text width=5.0em]{\small Velocity $\bu$}	
\addplot+[mark=triangle, mark size=2, color=blue, ultra thick] 
table[x expr=\thisrowno{0}, y expr=\thisrowno{1}, col sep=space] {Figures/MMF_structure.txt};
\addlegendentry[text width=5.0em]{\small Structure $\lambda$}	
\addplot[mark=none, very thick, dashed, color=black, domain=0.03:0.25, samples=100] {0.5*x^(3)}; 
\node[rotate=35]  at (axis cs:0.1, 0.0002)
{Referene slope 3};
\end{loglogaxis}
\end{tikzpicture}
\end{center}
\caption{
Convergence of numerical solution 
$(\bu^{h},\lambda^{h})$ compared with manufactured analytic solution
$(\widetilde{\bu},\widetilde{\lambda})$.
At each discretization with $N$-elements, 
the $L_2$-norm errors are plotted in log-log scale with 
the estimated cell average diameter 
$h_{\mathrm{est}}=1/\sqrt{N}$. 
Both numerical solution $\bu^{h},\lambda^{h}$
converges to the exact solution at 3rd order.}
\label{fig:mms}
\end{figure}

Finally, the numerical scheme described in Section~\ref{sec_Nscheme}
was used to solve the set of governing equation
\begin{equation}
\nabla \cdot \btau - \nabla p =\vec{F}, 
\end{equation}
\begin{equation}
\nabla \cdot \bu =0, 
\end{equation}
\begin{equation}
(\bu \cdot \nabla \lambda)+
k_d \dgamma \lambda - k_a (1-\lambda)=Q,
\end{equation}
and the order of accuracy is tested by comparing 
numerical solution $(\bu^{h},\lambda^{h})$ 
to $(\widetilde{\bu}, \widetilde{\lambda})$
at different number of discretizations. 
The scheme errors $\epsilon^{h}$ at each discretization 
are evaluated by the $L_2$-norm difference between 
$(\bu^{h},\lambda^{h})$ 
and $(\widetilde{\bu}, \widetilde{\lambda})$.
In Figure~\ref{fig:mms},
$\epsilon^{h}$ is plotted in log-log scale together 
with respect to $h_{\mathrm{est}}=1/\sqrt{N}$.
Here, $N$ is the total number of elements
(mesh cell).
In the main simulation conducted through out the paper, 
we used total $N=49,152$ mesh-elements, and
thus $h_{est}=0.0045$.
At a piecewise polynomial of order $d=2$ 
for a approximation function  
it was confirmed that errors decrease with the order 
of $d+1=3$ as theory of finite-element method suggests
\cite{Babuka:1982,Brenner}.
So, we conclude that our numerical scheme is 
verified.

\end{appendices}

\newpage
\bibliographystyle{unsrt}
\bibliography{paper_thixo}

\begin{thebibliography}{10}

\bibitem{Mewis:2011}
J.~Mewis and N.~J. Wagner.
\newblock {\em Colloidal Suspension Rheology}, chapter Thixotropy, pages
  228--251.
\newblock Cambridge Series in Chemical Engineering. Cambridge University Press,
  2011.

\bibitem{Larson:2019}
R.~G. Larson and Y.~Wei.
\newblock A review of thixotropy and its rheological modeling.
\newblock {\em Journal of Rheology}, 63:477--501, 2019.

\bibitem{Mewis:2006}
K.~Dullaert and J.~Mewis.
\newblock A structural kinetics model for thixotropy.
\newblock {\em Journal of Non-Newtonian Fluid Mechanics}, 139:21--30, 2006.

\bibitem{Food1}
J.~Engmann and A.~S. Burbidge.
\newblock Fluid mechanics of eating, swallowing and digestion - overview and
  perspectives.
\newblock {\em Food \& Function}, 4:443--447, 2013.

\bibitem{thixo_example_1}
D.~Quemada and R.~Droz.
\newblock Blood viscoelasticity and thixotropy from stress formation and
  relaxation measurements: a unified model.
\newblock {\em Biorheology}, 20:635--651, 1983.

\bibitem{thixo_example_2}
R.~G. de~Krester and D.~V. Boger.
\newblock A structural model for the time-dependent recovery of mineral
  suspensions.
\newblock {\em Rheologica Acta}, 40:582--590, 2001.

\bibitem{thixo_example_3}
Peptide-Based Physical Gels~Endowed with Thixotropic~Behaviour.
\newblock N. zanna and c. tomasini.
\newblock {\em Gels}, 3:39, 2017.

\bibitem{thixo_example_4}
S.~Mortazavi-Manesh and J.~M. Shaw.
\newblock Thixotropic rheological behavior of maya crude oil.
\newblock {\em Energy \& Fuels}, 28:972--979, 2014.

\bibitem{Armelin:2006}
E.~Armelin, M.~Mart{\'\i}, E.~Rud{\'e}, J.~Labanda, J.~Llorens, and
  C.~Alem{\'a}n.
\newblock A simple model to describe the thixotropic behavior of paints.
\newblock {\em Progress in Organic Coatings}, 57(3):229--235, 2006.

\bibitem{Barnes:1997}
H.~A. Barnes.
\newblock Thixotropy---a review.
\newblock {\em Journal of Non-Newtonian Fluid Mechanics}, 70(1--2):1--33, 1997.

\bibitem{Mewis:2009}
J.~Mewis and N.~J. Wagner.
\newblock Thixotropy.
\newblock {\em Advances in Colloid and Interface Science}, 147--148:214--227,
  2009.

\bibitem{Goodeve}
C.~F. Goodeve.
\newblock A general theory of thixotropy and viscosity.
\newblock {\em Transactions of Faraday Society}, 35:342--358, 1939.

\bibitem{Moore:1959}
F.~Moore.
\newblock The rheology of ceramic slips and bodies.
\newblock {\em Transactions and journal of British Ceramic Society},
  58:470--494, 1959.

\bibitem{Stickel:2006}
J.~Stickel, R.~J. Phillips, and R.~L Powell.
\newblock A constitutive model for microstructure and total stress in
  particulate suspensions.
\newblock {\em Journal of Rheology}, 50:379, 2006.

\bibitem{Goddard:1984}
J.~D. Goddard.
\newblock Dissipative materials as models of thixotropy and plasticity.
\newblock {\em Journal of Non-Newtonian Fluid Mechanics}, 14:141--160, 1984.

\bibitem{Patel:1988}
P.~D. Patel and W.~B. Russel.
\newblock A mean field theory for the rheology of phase separated or
  flocculated dispersions.
\newblock {\em Colloids and Surfaces}, 31:355--383, 1988.

\bibitem{Potanin:1991}
A.~A. Potanin.
\newblock On the mechanism of aggregation in the shear flow of suspensions.
\newblock {\em Journal of Colloid and Interface Science}, 145:140--157, 1991.

\bibitem{Dickey:2004}
D.~S. Dickey and J.~B. Fasano.
\newblock How geometry and viscosity influence mixing.
\newblock {\em Chemical Engineering}, 111:42--46, 2004.

\bibitem{Metzner:1957}
A.~B. Metzner and R.~E. Otto.
\newblock Agitation of non‐{N}ewtonian fluids.
\newblock {\em AIChE Journal}, 3:3--10, 1957.

\bibitem{new_cite1}
J.~E. L{\'o}pez-Aguilar, M.~F. Webster, H.~R. Tamaddon-Jahromi, and O.~Manero.
\newblock A comparative numerical study of time-dependent structured fluids in
  complex flows.
\newblock {\em Rheologica Acta}, 55:197--214, 2016.

\bibitem{new_cite2}
J.~E. L{\'o}pez-Aguilar, M.~F. Webster, H.~R. Tamaddon-Jahromi, and O.~Manero.
\newblock Numerical modelling of thixotropic and viscoelastoplastic materials
  in complex flows.
\newblock {\em Rheologica Acta}, 54:307--325, 2014.

\bibitem{new_cite3}
J.~E. L{\'o}pez-Aguilar, M.~F. Webster, H.~R. Tamaddon-Jahromi, and O.~Manero.
\newblock High-weissenberg predictions for micellar fluids in
  contraction--expansion flows.
\newblock {\em Journal of Non-Newtonian Fluid Mechanics}, 222:190--208, 2015.

\bibitem{new_cite4}
J.~E. L{\'o}pez-Aguilar, M.~F. Webster, H.~R. Tamaddon-Jahromi, and O.~Manero.
\newblock A new constitutive model for worm-like micellar systems -- numerical
  simulation of confined contraction--expansion flows.
\newblock {\em Journal of Non-Newtonian Fluid Mechanics}, 204:7--21, 2014.

\bibitem{Coussot:2014}
P.~Coussot.
\newblock Yield stress fluid flows: A review of experimental data.
\newblock {\em Journal of Non-Newtonian Fluid Mechanics}, 211:31--49, 2014.

\bibitem{Beris:1985}
A.~N. Beris, J.~A. Tsamopoulos, R.~C. Armstrong, and R.A. Brown.
\newblock Creeping motion of a sphere through a {B}ingham plastic.
\newblock {\em Journal of Fluid Mechanics}, 1588:219--244, 1985.

\bibitem{Mitsoulis:1997}
J.~Blackery and E.~Mitsoulis.
\newblock Creeping motion of a sphere in tubes filled with a {B}ingham plastic
  material.
\newblock {\em Journal of Non-Newtonian Fluid Mechanics}, 70:59--77, 1997.

\bibitem{Tripathi:1995}
A.~Tripathi and R.~P. Chhabra.
\newblock Drag on spheroidal particles in dilatant fluids.
\newblock {\em AIChE Journal}, 41:728--731, 1995.

\bibitem{Tian:2018}
S.~Tian.
\newblock Wall effect for spherical partical in confined shear-thickening
  fluids.
\newblock {\em Journal of Non-Newtonian Fluid Mechanics}, 257:13--21, 2018.

\bibitem{Gervang:1992}
B.~Gervang, A.~R. Davies, and T.~N. Phillips.
\newblock On the simulation of viscoelastic flow past a sphere using spectral
  methods.
\newblock {\em Journal of Non-Newtonian Fluid Mechanics}, 44:281--306, 1992.

\bibitem{Huang:1995}
P.~Y. Huang and J.~Feng.
\newblock Wall effects on the flow of viscoelastic fluids around a circular
  cylinder.
\newblock {\em Journal of Non-Newtonian Fluid Mechanics}, 60:179--198, 1995.

\bibitem{Ferrioir:2004}
T.~Ferroir, H.~Huynh, and X.~Chateau~P. Coussot.
\newblock Motion of a solid object through a pasty (thixotropic) fluid.
\newblock {\em Physics of Fluids}, 16(3):594--601, 2004.

\bibitem{Jirsaraei:2018}
N.~Maleki-Jirsaraei, S.~Hassani, and S.~Azizi.
\newblock Settling of spherical objects through thixotropic fluids: A
  statistical approach.
\newblock {\em Modern Applied Science}, 12:72--76, 2018.

\bibitem{Gumulya:2014}
M.~M. Gumulya, R.~R. Horsley, and V.~Pareek.
\newblock Numerical simulations of the settling behavior of particles in
  thixotropic fluids.
\newblock {\em Physics of Fluids}, 26, 2014.

\bibitem{Thant:2011}
M.~M.~M. Thant, M.~T.~M. Sallehud-Din, G.~Hewitt, C.~Hale, and J.~Quarini.
\newblock Mitigating flow assurance challenges in deepwater fields using active
  heating methods.
\newblock {\em Society of Petroleum Engineer}, 2011.

\bibitem{Huang:1976}
C.~R. Huang and W.~Fabisiak.
\newblock Thixotropic parameters of whole human blood.
\newblock {\em Thrombosis Research,}, 8:1--8, 1976.

\bibitem{Derksen:2010}
J.~J. Derksen.
\newblock Simulations of thixotropic liquids.
\newblock {\em Applied Mathematical Modeling}, 35(4):1656--1665, 2011.

\bibitem{Freund:2018}
J.~B. Freund, J.~Kim, and R.~H. Ewoldt.
\newblock Field sensitivity of flow predictions to rheological parameters.
\newblock {\em Journal of Non-Newtonian Fluid Mechanics}, 257, 2018.

\bibitem{Moore_example1}
C.~M. Rodkiewicz, editor.
\newblock {\em Arteries and Arterial Blood Flow: Biological and physiological
  aspects}.
\newblock CISM International Centre for Mechanical Sciences. Springer-Verlag
  Wien, 1st edition, 1983.

\bibitem{Moore_example2}
A.~Tehrani.
\newblock Thixotropy in water-based drilling fluids.
\newblock {\em Annual Transactions of the nordic rheology society}, 16:1--13,
  2008.

\bibitem{Happel:1965}
J.~Happel and H.~Brenner.
\newblock {\em Low Reynolds number hydrodynamics}.
\newblock Prentice-Hall, London, 1965.

\bibitem{Bruyn:2007}
H.~Tabuteau, P.~Coussot, and J.~de~Bruyn.
\newblock Drag force on a sphere in steady motion through a yield-stress fluid.
\newblock {\em Journal of Rheology}, 51(1):125--137, 2007.

\bibitem{Arndt:2017}
D.~Arndt, W.~Bangerth, D.~Davydov, T.~Heister, L.~Heltai, M.~Kronbichler,
  M.~Maier, J.~P. Pelteret, B.~Turcksin, and D.~Wells.
\newblock The \texttt{deal.II} library, version 8.5.
\newblock {\em Journal of Numerical Mathematics}, 2017.

\bibitem{Bangerth:2007}
W.~Bangerth, R.~Hartmann, and G.~Kanschat.
\newblock {deal.II} --- a general purpose object oriented finite element
  library.
\newblock {\em ACM Trans. Math. Softw.}, 33(4):24/1--24/7, 2007.

\bibitem{TaylorHood}
C.~Taylor and P.~Hood.
\newblock A numerical solution of the navier-stokes equations using the finite
  element technique.
\newblock {\em Computers \& Fluids}, 1:73--100, 1973.

\bibitem{DGbook}
J.~S. Hesthaven and T.~Warburton.
\newblock {\em Nodal Discontinuous {G}alerkin Methods: Algorithms, Analysis,
  and applications}.
\newblock Texts in Applied Mathematics. Springer-Verlag New York, 2008.

\bibitem{Saramito:2016_book}
P.~Saramito.
\newblock {\em Complex Fluids: Modeling and Algorithms}, volume~79 of {\em
  Math{\'e}matiques et Applications}.
\newblock Springer, 2016.

\bibitem{GMRES}
Y.~Saad and M.~H. Schultz.
\newblock {GMRES}: A generalized minimal residual algorithm for solving
  nonsymmetric linear systems.
\newblock {\em {SIAM} Journal on Scientific Computing}, 7:856--869, 1986.

\bibitem{ILU}
R.~C.~Mittal and. A.~H. AL-Kurdi.
\newblock An efficient method for constructing an ilu preconditioner for
  solving large sparse nonsymmetric linear systems by the gmres method.
\newblock {\em Computers \& Mathematics with applications}, 45:1757--1772,
  2003.

\bibitem{Bird:1987}
R.~B. Bird, R.~C. Armstrong, and O.~Hassager.
\newblock {\em Dynamics of {P}olymeric {L}iquids, \uppercase{V}ol. 1:
  \uppercase{F}luid mechanics}.
\newblock Wiley, 2nd edition, 1987.

\bibitem{Tsamopoulos:2016}
D.~Fraggedakis, Y.~Dimakopoulos, and J.~Tsamopoulos.
\newblock Yielding the yield-stress analysis: a study focused on the effects of
  elasticity on the settling of a single spherical particle in simple
  yield-stress fluids.
\newblock {\em Soft Matter}, 12(24):5378--5401, 2016.

\bibitem{Cross:1965}
M.~M. Cross.
\newblock Rheology of non-newtonian fluids: A new flow equation for
  pseudoplastic systems.
\newblock {\em Journal of Colloid Science}, 20:417--437, 1965.

\bibitem{Barnes:1999}
Howard~A. Barnes.
\newblock The yield stress---a review or ``panta rei ''--- everything flows?
\newblock {\em Journal of Non-Newtonian Fluid Mechanics}, 81(1):133 -- 178,
  1999.

\bibitem{JK:2019}
J.~Kim, P.~K. Singh, J.~B. Freund, and R.~H. Ewoldt.
\newblock Uncertainty propagation in simulation predictions of generalized
  newtonian fluid flows.
\newblock {\em Journal of Non-Newtonian Fluid Mechanics}, 271:104138, 2019.

\bibitem{Mitsoulis:2017}
E.~Mitsoulis and J.~Tsamopoulos.
\newblock Numerical simulations of complex yield-stress fluid flows.
\newblock {\em Rheologica Acta}, 56:231--258, 2017.

\bibitem{Balamforth:2014}
N.~J. Balmforth, I.~A. Frigaard, and G.~Ovarlez.
\newblock Yielding to stress: Recent developments in viscoplastic fluid
  mechanics.
\newblock {\em Annual Review of Fluid Mechanics}, 46:121--146, 2014.

\bibitem{Ardakani:2011}
H.~A. Ardakani, E.~Mitsoulis, and S.~G. Hatzikiriakos.
\newblock Thixotropic flow of toothpaste through extrusion dies.
\newblock {\em Journal of Non-Newtonian Fluid Mechanics}, 166:1262--1271, 2011.

\bibitem{Alexandrou:2009}
A.~N. Alexandrou, N.~Constantinou, and G.~Georgiou.
\newblock Shear rejuvenation, aging and shear banding in yield stress fluids.
\newblock {\em Journal of Non-Newtonian Fluid Mechanics}, 158:6--17, 2009.

\bibitem{Thompson:2018}
R.~L. Thompson, L.~U.~R. Sica, and P.~R.~S. Mendes.
\newblock The yield stress tensor.
\newblock {\em Journal of Non-Newtonian Fluid Mechanics}, 261:211--219, 2018.

\bibitem{Dinkgreve2017}
M.~Dinkgreve, M.~M. Denn, and D.~Bonn.
\newblock ``everything flows?'': elastic effects on startup flows of
  yield-stress fluids.
\newblock {\em Rheologica Acta}, 56(3):189--194, 2017.

\bibitem{new_cite5}
J.~E. L{\'o}pez-Aguilar, M.~F. Webster, H.~R. Tamaddon-Jahromi, and O.~Manero.
\newblock Predictions for circular contraction-expansion flows with
  viscoelastoplastic \& thixotropic fluids.
\newblock {\em Journal of Non-Newtonian Fluid Mechanics}, 261:188--210, 2018.

\bibitem{Renardy:2016}
M.~Renardy and Y.~Renardy.
\newblock Thixotropy in yield stress fluids as a limit of viscoelasticity.
\newblock {\em IMA Journal of Applied Mathematics}, 3:522--537, 2016.

\bibitem{Stephanou:2018}
P.~S Stephanou and G.~G. Georgiou.
\newblock A nonequilibrium thermodynamics perspective of thixotropy.
\newblock {\em The Journal of Chemical Physics}, 149:244902, 2018.

\bibitem{Lin:1985}
S.~F. Lin and R.~S. Brodkey.
\newblock Rheological properties of slurry fuels.
\newblock {\em Journal of Rheology}, 29(2):147--175, 1985.

\bibitem{Pinder:1964}
K.~L. Pinder.
\newblock Time dependent rheology of the tetrahydrofuran-hydrogen sulphide gas
  hydrate slurry.
\newblock {\em The Canadian Journal of Chemical Engineering}, 42:132--138,
  1964.

\bibitem{Burgos:2001}
G.~R. Burgos, A.~N. Alexandrou, and V.~Entov.
\newblock Thixotropic rheology of semisolid metal suspensions.
\newblock {\em Journal of Materials Processing Technology}, 110:164--176, 2001.

\bibitem{Freund:2015}
J.~B. Freund and R.~H. Ewoldt.
\newblock Quantitative rheological model selection: {G}ood fits versus credible
  models using {B}ayesian inference.
\newblock {\em Journal of Rheology}, 59:667--701, 2015.

\bibitem{Jeffreys:1939}
H.~Jeffreys.
\newblock {\em Theory of Probability}.
\newblock Oxford University Press, 3rd edition, 1961.

\bibitem{ftw_oblate}
X.~Yang, H.~Huang, and X.~Lu.
\newblock Sedimentation of an oblate ellipsoid in narrow tubes.
\newblock {\em Physical Review E}, 92:063009, 2015.

\bibitem{ftw_cylinder}
I.~E. Kareva and V.~L. Sennitskii.
\newblock Motion of a circular cylinder in a vibrating liquid.
\newblock {\em Journal of Applied Mechanics and Technical Physics},
  42:276--278, 2001.

\bibitem{ftw_cube}
N.~Agarwal and R.~P. Chhabra.
\newblock Settling velocity of cubes in newtonian and power law liquids.
\newblock {\em Powder Technology}, 178:17--21, 2007.

\bibitem{ftw_unsteady1}
T.~Sarpkaya.
\newblock Forces on cylinders and spheres in a sinusoidally oscillating fluid.
\newblock {\em Journal of Applied Mechanics}, 42:32--37, 1974.

\bibitem{ftw_unsteady2}
V.~L. Sennitskii.
\newblock Motion of a sphere in a vibrating liquid in the presence of a wall.
\newblock {\em Journal of Applied Mechanics and Technical Physics},
  40:662--668, 1999.

\bibitem{ftw_bubble}
L.~Zhang, C.~Yang, and Z-S. Mao.
\newblock Numerical simulation of a bubble rising in shear-thinning fluids.
\newblock {\em Journal of Non-Newtonian Fluid Mechanics}, 165:555--567, 2010.

\bibitem{assurance1}
W.~C. Chin.
\newblock {\em Computational Rheology for Pipeline and Annular Flow:
  Non-Newtonian Flow Modeling for Drilling and Production, and Flow Assurance
  Methods in Subsea Pipeline Design}.
\newblock Elsevier Science, 2001.

\bibitem{assurance2}
Ahmed J.~Ratulowski A.~Hammami.
\newblock {\em Precipitation and Deposition of Asphaltenes in Production
  Systems: A Flow Assurance Overview}, pages 617--660.
\newblock Springer New York, New York, NY, 2007.

\bibitem{Denn:2009}
D.~Bonn and M.~M. Denn.
\newblock Yield stress fluid slowly yield to analysis.
\newblock {\em Science}, 324:1401--1402, 2009.

\bibitem{Denn:2011}
M.~M. Denn and D.~Bonn.
\newblock Issues in the flow of yield-stress liquids.
\newblock {\em Rheologica Acta}, 50(4):307--315, 2011.

\bibitem{Mendes:2011}
P.~R.~S. Mendes.
\newblock Thixotropic elasto-viscoplastic model for structured fluids.
\newblock {\em Soft Matter}, 7(6):2471--2483, 2011.

\bibitem{MMS1}
C.~J. Roy.
\newblock Review of code and solution verification procedures for computational
  simulation.
\newblock {\em Journal of Computational Physics}, 205:131--156, 2005.

\bibitem{MMS2}
S.~Steinberg and P.~J. Roache.
\newblock Symbolic manipulation and computational fluid dynamics.
\newblock {\em Journal of Computational Physics}, 57:251--284, 1985.

\bibitem{MMS3}
P.~J. Roache.
\newblock Code verification by the {M}ethod of {M}anufactured solutions.
\newblock {\em Journal of Fluids Engineering}, 124:4--10, 2001.

\bibitem{Babuka:1982}
I.~Babuska and S.~Szabo.
\newblock On the rates of convergence of the finite element method.
\newblock {\em International Journal for Numerical Method for Engineering},
  18:323--341, 1982.

\bibitem{Brenner}
S.~Brenner and L.~R. Scott.
\newblock {\em The mathematical theory of finite element method}, volume~15 of
  {\em Texts in Applied Mathematics}.
\newblock Springer-Verlag New York, 2002.

\end{thebibliography}

\end{document}